\documentclass[aps, prb, twocolumn, superscriptaddress, english]{revtex4-2}

\usepackage{graphicx}
\usepackage{cancel}
\usepackage{enumerate}
\usepackage{hyperref}
\hypersetup{
colorlinks=true,
citecolor=blue,
linkcolor=blue,
urlcolor=blue,
}
\usepackage{textcomp}
\usepackage{amsmath}
\usepackage{amssymb}
\usepackage{pgf}
\usepackage{soul}
\usepackage{subfigure}
\usepackage[english]{babel}
\usepackage[normalem]{ulem}
\usepackage{upgreek}

\def\bra#1{\langle#1|}
\def\ket#1{|#1\rangle}

\def\ee{\mathrm{e}}

\newcommand{\Up}{{\uparrow}}
\newcommand{\Dn}{{\downarrow}}
\newcommand{\Upt}{{{\sf u}}}
\newcommand{\Dnt}{{{\sf d}}}
\newcommand{\Id}{{\mathbb{I}    }}

\begin{document}
\title{Cooper pair charge Kondo effect in a hybrid quantum dot - superconductor device }

\author{Luca Gonfiantini}

\affiliation{SISSA and INFN Sezione di Trieste, via Bonomea 265, 34136 Trieste, Italy}

\author{Lorenzo Maffi}

\affiliation{Dipartimento di Fisica e Astronomia “G. Galilei”,
Universit\`a degli Studi di Padova, I-35131 Padova, Italy}
\affiliation{Istituto Nazionale di Fisica Nucleare (INFN), Sezione di Padova, I-35131 Padova, Italy}

\author{Michele Burrello}

\affiliation{Dipartimento di Fisica dell’Università di Pisa and INFN, sezione di Pisa, Largo Pontecorvo 3, I-56127 Pisa, Italy}

\begin{abstract}

We consider a tunable hybrid device composed of four quantum dots and a floating superconducting island -- the so-called poor man's tetron -- close to its effective particle-hole symmetric point and coupled to four external leads. We predict the onset of a charge Kondo effect mediated by Cooper pairs, resulting from the strong crossed Andreev reflections that dominate its low-energy behavior. This two-channel Kondo effect makes the poor man's tetron behave as a double Cooper pair splitter, turning uncorrelated incoming electrons into spatially separated Cooper pairs. Based on renormalization group techniques, we analyze both its weak and strong coupling regimes and we determine its non-Fermi liquid signatures. Our estimates indicate that the fractional conductance and transport features associated to this two-channel charge Kondo effect can be observed over a broad range of temperatures and anisotropies in realistic experimental conditions.

\end{abstract}

\maketitle

\section{Introduction}\label{sec:intro}

The solution of the Kondo problem and the study of quantum impurity models are  milestones of modern condensed matter physics \cite{Hewson_1993}. They not only constituted a fundamental pillar for the modeling of strongly correlated materials \cite{Coleman_2007}, but also represented a crucial playground for the development of the theory of zero-dimensional non-Fermi liquids and one-dimensional integrable theories with the introduction of multi-channel Kondo models \cite{Affleck1990,Affleck1991}.  

\begin{figure} 
\includegraphics[width=\columnwidth]{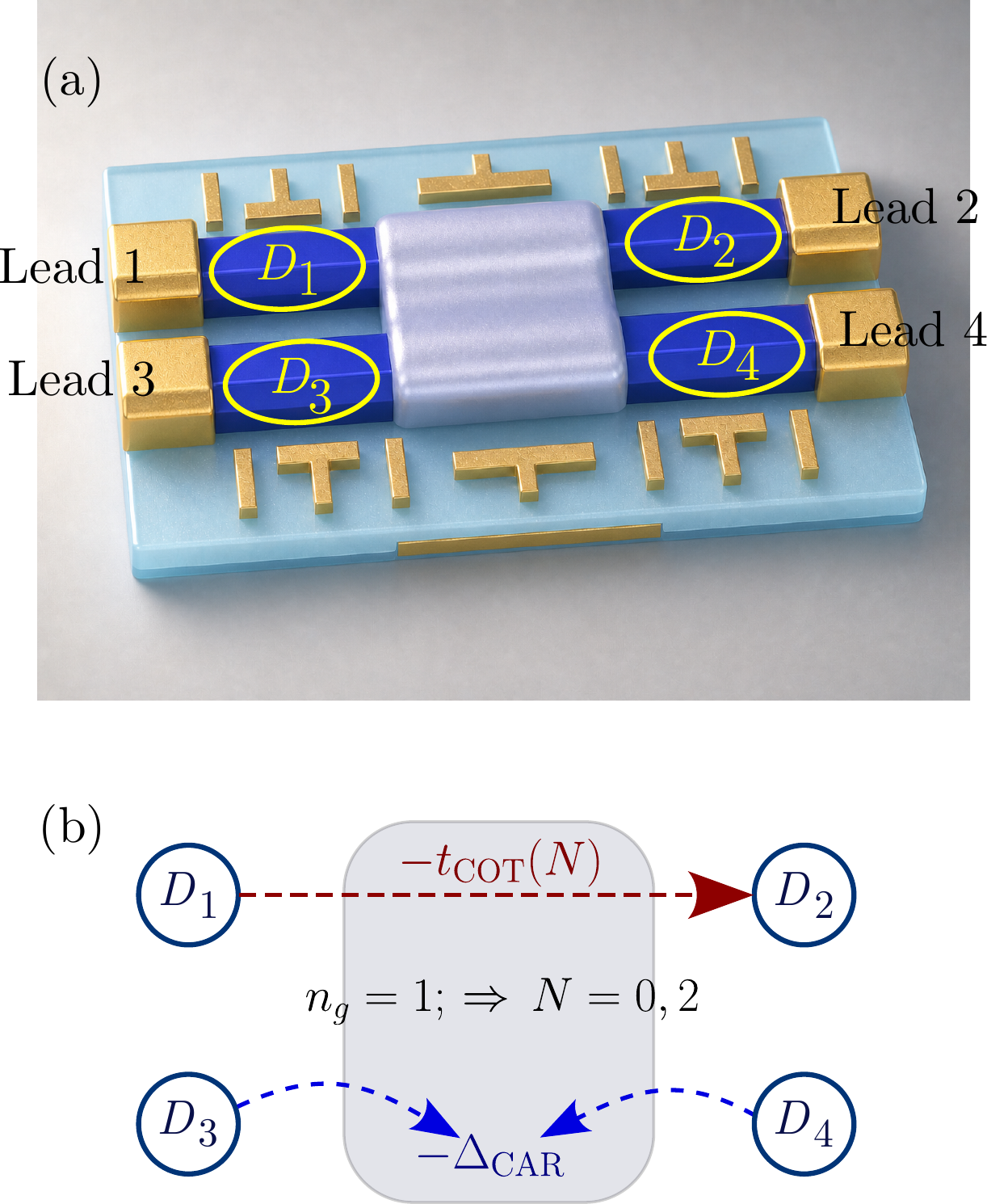}
\caption{Sketch of the poor man's tetron. (a) The four quantum dots $D_\alpha$ are realized within two parallel nanowires (blue) through suitable electrostatic gates, depicted in gold alongside them. The two nanowires are connected by a central floating superconducting island (gray). Each quantum dot is tunnel-coupled to the central superconducting region and an external lead. (b) Graphical representation of the cotunneling (red) and crossed Andreev reflection (blue) processes dictating the dynamics of electrons at low energy close to the charge degeneracy point.} \label{fig1}
\end{figure}

The two-channel Kondo effect provides indeed a textbook example of a non-Fermi liquid that displays a strong-coupling low-temperature critical point characterized by fractional conductance, anomalous critical exponents and a fractional zero-temperature residual entropy, which is the signature of an emerging decoupled Majorana degree of freedom \cite{EmeryKivelson,Mebrahtu2013}. These properties brought multi-channel Kondo models at the center of intense research for their implications in the study of heavy-electron materials \cite{Cox1987} and, more recently, for their relationship with anyonic zero-energy modes \cite{Lopes2020,Lotem2022}.

From the experimental point of view, the first observations of the two-channel Kondo model in controllable setups were obtained in double quantum dot systems \cite{Oreg2003,Potok2007}. 

More recently, another class of two-channel impurity models has been investigated based on the so-called charge Kondo effect, where the occupation of a quantum impurity plays the role of an effective spin degree of freedom \cite{Glazman1990,Matveev1991,Matveev1995}. This opened the path for the observation of multi-channel Kondo effects in quantum Hall systems \cite{Iftikhar2015,Iftikhar2018}, with striking confirmations of their predicted universal behavior \cite{Pustilnik2004,Mitchell2016}.

In this work we investigate the onset of a two-channel charge Kondo effect in a hybrid quantum dot - superconductor interacting device. We study, in particular, the charge Kondo effect emerging from the crossed Andreev reflection processes in a system composed of four quantum dots connected by a floating superconducting island, the so-called poor man's tetron \cite{Nitsch2025}.

Our model displays critical and transport features analogous to other realizations of the charge Kondo effect mediated by Cooper pairs \cite{Garate2011,Pustilnik2017,Papaj2019}. Differently from former proposals, however,
the two-channel charge Kondo effect we analyze is obtained as an effect of crossed, rather than standard, Andreev reflections, thus by pairing processes involving electrons in different quantum dots. This implies that the involved degrees of freedom we exploit are exclusively spatial, such that the poor man's tetron in the two-channel charge Kondo regime can be interpreted as a double Cooper pair splitter that collects uncorrelated electrons from two source leads, and returns correlated electrons emitted by a crossed Andreev process into two distinct drain leads.

Analogously to the proposal in Ref. \cite{Pustilnik2017}, the device that we study is inspired by recent experiments in hybrid semiconductor-superconductor systems that provide a high tunability of all their physical parameters based on electrostatic gates. These devices have been successfully used for the implementation of Cooper pair splitters with triplet pairing \cite{Wang2022,Wang2023} and poor man's Majorana modes \cite{Bordin2023,Dvir2023,Zatelli2023,bordin2024,haaf2024,haaf2025,kulesh2025,vanLoo2026}.

In order to extend their use for the implementation of strongly-correlated systems, however, we consider setups with floating superconducting islands coupled to external dots \cite{SoutoBaran,Nitsch2025}. In particular, our construction relies on double-nanowire architectures which can be implemented either based on pairs of in situ grown parallel nanowires 
\cite{Kurtossy2021,Kanne2022,Vekris2022,Kanne2022,Kurtossy2026} or via quantum dots and nanowires electrostatically defined in 2D electron gases \cite{Wang2023,haaf2024,haaf2025,kulesh2025} (see, for instance the two-nanowire tetron device in Ref. \cite{Microsoft2025b}).

These hybrid semiconductor - superconductor platforms pose the basis of the poor man's tetron device, which is introduced in Sec. \ref{sec:PMT}. There, we present its low-energy description, instrumental to define the effective two-level system that we adopt for investigating the Cooper pair charge Kondo effect. In particular, the ground state degeneracy, necessary for the onset of the Kondo effect, is guaranteed by the presence of a particle-hole symmetry of the poor man's tetron Hamiltonian for specific values of its parameters. Sec. \ref{sec:Kondo} is devoted to the definition of the Kondo model we address, based on the ground states of the poor man's tetron tuned at the particle-hole symmetric point. In Sec. \ref{sec:weak}, we perform a renormalization group (RG) analysis of the Kondo Hamiltonian in a weak coupling regime. We show that, for a suitable choice of the initial couplings, the RG flow drives the system toward a two-channel Kondo fixed point at strong coupling, signaling the emergence of a non-Fermi liquid behavior. We conclude this section by characterizing, at weak coupling, the charge transport across the device, mediated by crossed Andreev reflection processes. In Sec. \ref{sec:strong}, we turn to a strong coupling analysis. This allows us to characterize both the universal conductance at the two-channel Kondo fixed point and the finite temperature corrections away from it. Finally, in Sec. \ref{sec:experiments}, we give a detailed account of possible perturbations to the charge Kondo Hamiltonian. We find that the non-Fermi liquid signatures of the two-channel Kondo fixed point can be observed experimentally over a wide range of parameters. Technical details of the low-energy description of both the poor man's tetron and the charge Kondo Hamiltonians are provided in the appendices, which additionally include a detailed discussion of the RG analysis at both weak and  strong coupling.

\section{The poor man's tetron in the charge degeneracy regime} \label{sec:PMT}

\subsection{The poor man's tetron}

The theoretical design of the poor man's tetron, presented in Ref. \cite{Nitsch2025}, has been devised to obtain a tunable platform for the study of exotic Kondo effects. The poor man's tetron is a device composed of four quantum dots connected by a central superconducting region. The dots are defined on a pair of parallel nanowires through suitable electrostatic gates. We assume that their electrostatic interaction is strong, such that each quantum dot displays only one orbital relevant for the low-energy dynamics of the system. As depicted in Fig. \ref{fig1} (a), both nanowires are covered by the same floating superconducting island, such that we distinguish two left quantum dots ($\eta=L$) and two right quantum dots ($\eta = R$). The central segment of each nanowire is affected by a proximity-induced pairing interaction and can be approximately described as a further central hybrid semiconductor-superconductor quantum dot \cite{SoutoBaran}, with the scope of modeling the behavior of a subgap Andreev state. These Andreev states are necessary to obtain the cotunneling and crossed Andreev processes \cite{Souto_arXiv2024} that constitute the main interactions connecting the two external quantum dots of each nanowire at low energy \cite{Liu2022,Bordin2023} [Fig. \ref{fig1} (b)].

 In the presence of a strong out-of-plane magnetic field, the four external dots become spin-polarized as an effect of the Zeeman interaction, whereas we assume that the central hybrid dots are screened by the overlaying superconducting island. To describe this strongly polarized regime for the external dots, we introduce the annihilation and creation operators $d_{\tau\eta}$ and $d^\dag_{\tau\eta}$ of the electrons in the dot of the $\tau$ nanowire ($\tau=\Upt,\Dnt$) on the $\eta$ side. We also adopt operators $c_{\tau s}$ and $c^\dag_{\tau s}$ to refer to the electron in the central superconducting region of the nanowire $\tau$, with spin $s$. The single-particle dynamics of the model can thus be approximated by the Hamiltonian: 
\begin{multline} \label{hamsp}
    H_{\text{sp}}=\mu_D\sum_{\tau,\eta}d^{\dagger}_{\tau\eta}d_{\tau\eta} +\sum_{\tau,s}\mu_{SC}c^{\dagger}_{\tau s}c_{\tau s}\\
    +\Delta\sum_{\tau}\left(c^{\dagger}_{\tau \uparrow }c^{\dagger}_{\tau \downarrow }\ee^{i\varphi}+ {\rm H.c.}\right) \\ -t \sum_{\tau,\eta}\left[\cos \alpha \, c^{\dagger}_{\tau \Dn} d_{\tau \eta } - \eta \sin \alpha \, c^{\dagger}_{\tau \Up} d_{\tau \eta }  + {\rm H.c.} \right].
\end{multline}
In this equation, $\mu_D$ represents the energy level of the four external quantum dots (which depends on the dot's electrostatic interactions, induced charge and Zeeman energy), whereas $t$ labels the tunneling amplitude from the external dots to the central superconducting region.  as an effect of the spin-orbit coupling of the semiconductor nanowires, the electron spin is rotated during the tunneling events; the parameter $\alpha$ is proportional to an effective Rashba spin-orbit momentum and the tunneling choice reflects a spin rotation around the $\sigma_y$ axis, thus orthogonal to the out-of-plane direction which is expected to approximately identify the gradient of the chemical potential of the heterostructure. In Eq. \eqref{hamsp}, $\eta=\pm 1$ for left and right dots respectively; this corresponds to the same spin rotation for hopping events in the same direction. We assume for simplicity that the previous parameters are tuned to the same value for all quantum dots, although this assumption is not necessary for our analysis (see Appendix \ref{app:pert} for more detail). Finally, $\Delta$ represents the amplitude of the proximity-induced s-wave pairing in the central superconducting regions, whereas $\ee^{i\varphi}$ constitutes a phase operator that describes the annihilation of a Cooper pair in the superconducting island. We additionally define a related number operator $N_{\rm CP}$ for the Cooper pairs, such that $\left[N_{\rm CP},\ee^{i\varphi}\right]=-\ee^{i\varphi}$, which is instrumental to define the electrostatic energy of the superconducting island as:
\begin{equation} \label{charging}
H_{\rm c} = E_C \left(N-n_g\right)^2\,.
\end{equation} 
Here the operator
\begin{equation}
N=2N_{\rm CP} + \sum_{\tau,s} c^\dag_{\tau s} c_{\tau s}
\end{equation}
represents the excess number of electrons of the superconducting region, whereas the parameter $n_g$ is the charge induced in the superconducting island by the external potentials and can be controlled through suitable electrostatic gates. The Hamiltonian $H_{\rm c}$ provides a simple description for the electrostatic interactions caused by the floating superconducting island where $E_C$ represents the typical charging energy scale. More complex electrostatic interactions, involving also the dot charges, can be considered without hindering the  analysis in the following sections.

The Hamiltonian $H_{\rm PMT}= H_{\rm sp} + H_{\rm c}$ constitutes the starting point for the definition of the effective quantum impurity that we will adopt to study the Cooper-pair charge Kondo effect, and it is characterized by a total conserved charge,
\begin{equation}
N_t= \sum_{\tau,\eta} d^\dag_{\tau \eta} d_{\tau \eta} + N\,,
\end{equation}
which we will use to characterize its eigenstates.
Additionally, Eq. \eqref{hamsp} describes a system where any single-electron tunneling from one nanowire and the other is absent. This assumption corresponds to devices in which only the bulk of the superconducting island connects the two nanowires, such that both the bulk superconducting gap and the disorder characterizing the superconducting island strongly suppress the exchange of quasi-particles between the nanowires at low energies. Due to this assumption it is easy to verify that the fermionic parities $P_\tau = (-1)^{\sum_\eta d^\dag_{\tau \eta}d_{\tau \eta} + \sum_s c^\dag_{\tau s} c_{\tau s}} = \pm 1$ are conserved quantities.

To gain insight on the low-energy behavior of the poor man's tetron, it is useful to consider the weak-tunneling limit $t \ll \Delta - E_C$ where the dynamics of the system can be described with a second-order perturbation theory in terms of elastic cotunneling and crossed Andreev reflection processes \cite{SoutoBaran,Nitsch2025}:
\begin{multline} \label{hameff}
H_{\rm eff} = \\
-\left[t_{\rm COT}(N) \sum_{\tau} d^\dag_{\tau R} d_{\tau L} +\Delta_{\rm CAR} \ee^{-i\varphi} \sum_{\tau} d_{\tau R} d_{\tau L} + {\rm H.c.}\right] 
\\ +\mu_{D}  \sum_{\tau,\eta} d^\dag_{\tau\eta}d_{\tau\eta} + E_C \left(N-n_g\right)^2 + H_{\rm SHIFT}(N) \,.
\end{multline}
Elastic cotunneling processes correspond to the hopping of an electron from one quantum dot to the other within the same nanowire [Fig. \ref{fig1}(b)]. The derivation of Eq. \eqref{hameff} (see Appendix \ref{app:pert}) shows that the related cotunneling amplitude $t_{\rm COT}$ depends on the excess charge $N$ of the superconducting island. 

Crossed Andreev reflections, instead, describe second-order processes in which an electron from one quantum dot pairs with an electron from the other quantum dot on the same nanowire and they create a Cooper pair [Fig. \ref{fig1}(b)]. As a result, the operator $\ee^{-i\varphi}$ increases the island charge $N$ by 2 electrons, $\left[N,\ee^{-i\varphi} \right]=2\ee^{-i\varphi}$. The crossed Andreev amplitude $\Delta_{\rm CAR}$, differently from the cotunneling processes, is independent of the charge $N$. A rigorous perturbative definition of the crossed Andreev processes at second order requires that $\Delta \gg t \gtrsim E_c$ (see Appendix \ref{app:pert}); we expect, however, that the emergence of the Cooper pair charge Kondo effect is favored by large tunneling rates $t \sim \Delta$, which are expected to yield larger energy gaps. We observe, in particular, that the superconducting pairing between the dots grows with $t$ and is more and more significant in the strong-tunneling regime, as demonstrated by recent experiments with quantum dots coupled via grounded superconducting elements \cite{Zatelli2023}.

Finally, Eq. \eqref{hameff} accounts for additional second-order processes where electrons tunnel back and forth between the same quantum dot and the superconducting region determine a charge-dependent shift of the quantum dot energy levels proportional to $t_{\rm COT}(N)/\cos 2\alpha$. These corrections to the dot potential and the charging energy interaction are captured by the term $H_{\rm SHIFT}(N)$ (see Appendix \ref{app:pert}).

As in the case of $H_{\rm PMT}$, also $H_{\rm eff}$ commutes with the total charge $N_t$ and the fermionic parities $P_\tau$, which, in the perturbative limit, depend on the dot occupation numbers only, as the occupation of the quasiparticle excitations of the superconducting island is suppressed by the superconducting gap $\Delta$. Therefore, in the following, we will analyze the poor man's tetron states based on a decomposition of the Hamiltonian in sectors labelled by the quantum numbers $\left(N_t,P_\Dnt\right)$, as the parity $P_\Upt=(-1)^{N_t}P_\Dnt$ is not independent of the other two conserved quantities. The analysis of the effective Hamiltonian blocks in each symmetry sector is presented in Appendix \ref{app:sectors}.

\subsection{Charge degeneracy conditions and symmetry sectors}

Charge Kondo effects arise from the degeneracy of low-energy states with different charge, as in the simplest case of a spin-polarized quantum dot tuned at its charge degeneracy point.

In the case of the poor man's tetron, we aim at identifying specific conditions under which the ground states of Hilbert space sectors with different even total charge, for instance $N_t=2$ and $N_t=4$, become degenerate. 

The Hamiltonian $H_{\rm PMT}$ is periodic in the induced charge $n_g$ with a period of 2 electrons: therefore, we focus without loss of generality on the interval $n_g \in [0,2]$. Consequently, in the regime with strong interactions, $E_C \gg \left|t_{\rm COT}\right|, \Delta_{\rm CAR}$, the charging energy contribution to the Hamiltonian causes the lowest energy sectors to be the ones with total charge $N_t=2,3,4$.

Driven by the analogy with the charge degeneracy point of a single quantum dot, we take advantage of the emergence of a particle-hole (PH) symmetry of the Hamiltonian $H_{\rm PMT}$ at specific values of its parameters to identify conditions under which the sectors with $N_t=2$ and $N_t=4$ become degenerate. For the onset of a Cooper pair charge Kondo effect, however, the exact PH symmetry is not required. The degeneracy of the ground states of the $N_t=2$ and $N_t=4$ sectors is indeed sufficient. The PH symmetry is useful to flag sufficient conditions for this degeneracy, but, if the system is perturbed around the PH symmetric point, the phase boundaries separating the $N_t=2$ and $N_t=4$ ground state regions retain the required ground state degeneracy.

The PH symmetry corresponds to the following mapping between the operators in Eqs. (\ref{hamsp},\ref{charging}): 
\begin{align}
&d_{\tau\eta} \leftrightarrow d^\dag_{\tau\eta}\,, \qquad c_{\tau s} \leftrightarrow -c^\dag_{\tau s} \label{PH1}\\
&\ee^{i\varphi} \leftrightarrow -\ee^{-i \varphi}\,, \quad N_t \leftrightarrow 4+2n_g-N_t \,, \label{PH2}
\end{align}
where the last equation also implies $N \leftrightarrow 2n_g -N$ and matches the expected PH symmetry of the charging energy $H_c$. The PH relations for $N_t$ and $N$ restrict the possible values of the induced $n_g$ to integers or half-integer numbers. As we wish to address a Cooper-pair charge Kondo effect in a superconducting system, we restrict to the case of integer $n_g$ and we set $n_g=1$. Finally, it is easy to verify that the Hamiltonian $H_{\rm sp}$ is PH symmetric under the transformation in Eq. \eqref{PH1} if and only if $\mu_D=\mu_{SC}=0$. These conditions, in particular, yield a two by two degeneracy of all the eigenstates in the sectors with $N_t=2$ and $N_t=4$. This degeneracy is stable against local perturbations of the $t$ and $\alpha$ parameters, such that it does not depend on spatial symmetries and is robust against anisotropies of the tunneling terms.

To understand the behavior under the PH symmetry of the sectors with $N_t=2$ and $4$, it is instructive to consider the effective Hamiltonian $H_{\rm eff}$, which inherits the same symmetry from $H_{\rm PMT}$. The mapping between different charge sectors of the Hamiltonian $H_{\rm eff}$ is obtained by considering the different behavior under the charge conjugation transformations in Eqs. \eqref{PH1} and \eqref{PH2} of the crossed Andreev and cotunneling amplitudes. The Andreev amplitude $\Delta_{\rm CAR}$ does not depend on the charge $N$, therefore its value is unchanged under PH symmetry. In particular, for $\mu_{SC}=0$, the amplitude reads (see Appendix \ref{app:pert}):
\begin{equation} \label{deltacar}
\Delta_{\rm CAR} = \frac{t^2 \sin\left(2\alpha\right)}{{\Delta}-E_C}\,.
\end{equation}
The situation is different for the amplitude $t_{\rm COT}$ and the corrections to the quantum dot potentials within $H_{\rm SHIFT}$. These parameters depend indeed on the charge $N$ and, at the charge degeneracy point ($n_g=1$, $\mu_D=\mu_{SC}=0$), we obtain $t_{\rm COT}(N)=-t_{\rm COT}(2-N)$ (see Appendix \ref{app:pert}). This transformation concurs to the exact mapping between the sectors $(N_t=2,P_\Dnt)$ and $(N_t=4,P_\Dnt)$, which offer the kind of ground state degeneracy needed for the onset of a Cooper pair charge Kondo effect. The related cotunneling amplitudes result:
\begin{equation} \label{tcot}
t_{\rm COT}(N=0) = - t_{\rm COT}(N=2) = \frac{2t^2E_C\cos(2\alpha)}{\left(\Delta+3E_C\right)\left(\Delta-E_C\right)}.
\end{equation}

\begin{figure}[t]
\includegraphics[width=\columnwidth]{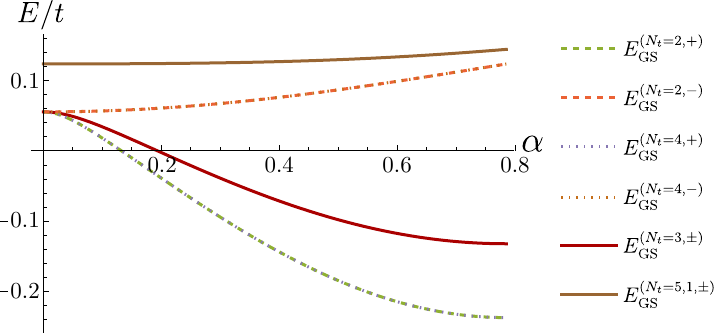}
\caption{Energy of the ground states of the sectors with $1\le N_t \le 5$ as a function of the spin-orbit parameter $\alpha$ at the charge degeneracy point $n_g=1$, $\mu_D=\mu_{SC}=0$, for $\Delta=5t$ and $E_C=0.5 t$. The energies are calculated based on the effective Hamiltonian \eqref{hameff} and expressed in units of the tunneling amplitude $t$. The PH symmetric ground states with $N_t=2,4$ and $P_\Dnt=+1$ are the global ground states for $\alpha>0$. At $\alpha=\pi/6$, $t_{\rm COT}(0)\approx 0.02t$ and $\Delta_{\rm CAR} \approx 0.19 t$.} \label{fig:GSenergy}
\end{figure}

 Within the effective Hamiltonian approximations $(\Delta \gg t \gtrsim E_C)$ at the charge degeneracy point, we observe that for any value of $\alpha>0$ and $E_C<\Delta$, the degenerate ground states of the system belong to the degenerate $(2,+)$ and $(4,+)$ sectors, with the odd $N_t=3$ sector ground states acquiring an intermediate energy between the even sectors with $P_\Dnt=+1$ and $P_\Dnt=-1$ (see Appendix \ref{app:sectors} and Fig. \ref{fig:GSenergy}).

This hierarchy among the sector energies reflects the peculiarities of the system at the charge degeneracy point: the ratio between cotunneling and crossed Andreev reflection amplitudes results $t_{\rm COT}(0)/\Delta_{\rm CAR}= \frac{2E_C\cot{2\alpha}}{\Delta+3E_C}$; therefore Andreev processes dominate over cotunneling for a broad range of $\alpha$ if $E_C \ll \Delta$. This is the main difference between our regime of interest at $\mu_{\rm SC}=0$ and the ranges of parameters investigated for the search of poor man's Majorana modes in floating superconducting islands \cite{SoutoBaran,Nitsch2025}, which require similar values of $\Delta_{\rm CAR}$ and $\left| t_{\rm COT}\right|$. In particular, we observe that in the limit of vanishing charging energy $E_C$, the poor man's tetron at the charge degeneracy point is characterized by a vanishing cotunneling amplitude. In this limit, therefore, the system does not display the physics of poor man's Majorana modes in non-interacting devices \cite{Liu2022}, but rather it approaches a scenario with two uncorrelated Cooper pair splitters \cite{Recher2001,Hofstetter2009}, one for each nanowire.

\section{The Cooper pair charge Kondo model} \label{sec:Kondo}

The previous analysis shows that the poor man's tetron at the charge degeneracy point displays two degenerate ground states with $N_t=2,4$, separated by an energy gap from the ground states of the sectors with $N_t=3$. We label this gap by $\delta E^{(3)}$ [see Eq. \eqref{dE3}]; in a regime with $\Delta \gg E_C$, such that $\Delta_{\rm CAR} \gg t_{\rm COT}(0)$, a rough approximation of this gap is given by $\delta E^{(3)} \approx \left(\sqrt{2}-1 \right) \Delta_{\rm CAR}$.

In the following, we label the degenerate ground states of the sectors $(4,+)$ and $(2,+)$ respectively as $\ket{4}$ and $\ket{2}$ (see Appendix \ref{App:evenp} for more detail). For energy scales considerably below $\delta E^{(3)}$, the poor man's tetron at the PH symmetric point can thus be represented by an effective spin 1/2 degree of freedom determined by these two states. We accordingly introduce Pauli matrices $S_i$ such that $S^z \ket{4}=\ket{4}$ and $S^z \ket{2}=-\ket{2}$.
This is the degree of freedom which we adopt to set up our construction for the study of a two-channel charge Kondo effect. In particular, $S^z$ constitutes a charge degree of freedom ($N_t=4,2$) such that the processes described by $S^{\pm}=\left(S^x \pm iS^y\right)/2$ are equivalent to a Cooper pair moving  from the external environment to the poor man's tetron or vice versa.

As depicted in Fig. \ref{fig1}, we consider an environment composed of four external leads, each of which is coupled to one of the poor man's tetron's quantum dots.

\begin{figure}
\includegraphics[width=\columnwidth]{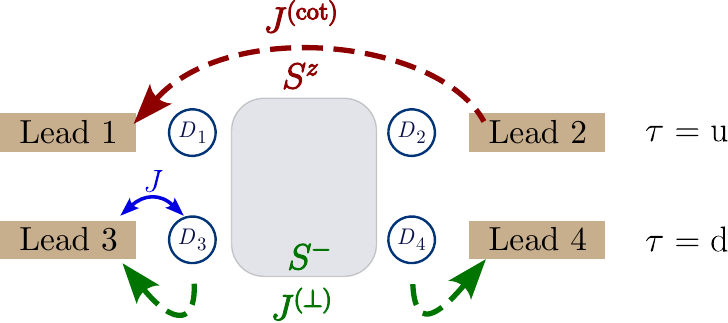}
\caption{Main processes in the Hamiltonian $H_K$. The blue arrow represents the (first-order) lead-dot couplings $H_J$. The dashed red and green arrows respectively depict the $J^{(\rm cot)}$ and $J^{(\perp)}$ second-order terms in Eqs. \eqref{Kondo2} and \eqref{Kondo3}.} \label{fig:Kondo}
\end{figure}

We consider therefore a global Hamiltonian of the following kind:
\begin{equation}
H_{\rm tot} = H_{\rm PMT} + H_{\rm leads} + H_J\,.
\end{equation}
The Hamiltonian $H_{\rm leads}$ defines the four leads as non-interacting one-dimensional models, whose energy levels are eventually set by external voltage biases. The tunneling interaction $H_J$ couples the leads and the poor man's tetron (see Fig. \ref{fig:Kondo}):
\begin{equation} \label{Hcoupling}
H_J= - \sum_{\alpha=1}^4 J_\alpha l^\dag_\alpha d_\alpha + {\rm H.c.}\,.
\end{equation}
For simplicity we labelled leads and quantum dots with the index $\alpha$ that runs from $1$ to $4$ and corresponds to the combinations of the indices $\tau$ and $\eta$ ordered as in Fig. \ref{fig1}. The operator $l^\dag_\alpha$ is the creation operator of an electron at the boundary of the lead $\alpha$. The amplitudes $J_\alpha$ can be chosen to be real and establish the coupling strength of the problem, and they can be experimentally controlled through suitable electrostatic cutter gates.

At low energies, the Hamiltonian $H_{\rm tot}$ identifies a strongly correlated system as an effect of the charging energy of the central superconducting region. To describe the low-energy dynamics of the system, we consider the weak coupling regime, identified by the conditions $J_\alpha \ll \delta E^{(3)}$. Under this assumption, the system can be effectively described by a second-order perturbative expansion in which the poor man's tetron is modelled as the two-level system defined by $\ket{4}$ and $\ket{2}$. We obtain the following effective Hamiltonian:
\begin{multline} \label{Kondo0}
H_{\rm K} = \sum_{\alpha \beta} \left( J_\alpha J_\beta l^\dag_\alpha l^\dag_\beta \Theta^{\alpha \beta} + {\rm H.c.}\right)\\
+ \sum_{\alpha \beta} J_\alpha J_\beta \left(  l_\beta l^\dag_\alpha  \Upsilon^{\alpha \beta} +  l^\dag_\alpha l_\beta \Xi^{\alpha \beta} \right)\,,
\end{multline}
where $\Theta$, $\Upsilon$ and $\Xi$ are $4\times 4$ matrices of operators acting on the effective spin-1/2 degree of freedom. In particular $\Theta\propto S^{-}$ describes pairs of electrons leaving the poor man's tetron, whereas $\Upsilon$ and $\Xi$ are respectively proportional to the projectors over $\ket{4}$ and $\ket{2}$, and correspond to processes in which one electron enters the device from lead $\beta$ and leaves it by moving into lead $\alpha$:
\begin{align}
&\Theta^{\alpha\beta}=\sum_{\substack{a=\pm \\n=1,.4}}-\frac{|2\rangle\langle2|d_{\alpha}|3,a,n\rangle\langle3,a,n|d_{\beta}|4\rangle\langle4|}{E_{n}^{(3)}-E_{GS}}\,, \label{Theta1}\\
&\Upsilon^{\alpha\beta}=\sum_{\substack{a=\pm \\n=1,.4}}-\frac{|4\rangle\langle4|d^{\dagger }_{\beta}|3,a,n\rangle\langle3,a,n|d_{\alpha}|4\rangle\langle4|}{E_{n}^{(3)}-E_{GS}} \,, \label{Upsilon1}\\
&\Xi^{\alpha\beta}=
     \sum_{\substack{a=\pm \\n=1,.4}}-\frac{|2\rangle\langle2|d_{\alpha}|3,a,n\rangle\langle3,a,n|d^{\dagger}_{\beta}|2\rangle\langle2|}{E_{n}^{(3)}-E_{GS}}\label{Xi1}\,.
\end{align}
In these expressions, the index $a$ labels the parity $P_\Dnt$ of the $N_t=3$ sectors, $n$ their four eigenstates and $E_{n}^{(3)}$ the related energies. $E_{GS}$ is the energy of the global ground states $\ket{4}$ and $\ket{2}$ such that, for $n=1,2$, $E_{n}^{(3)}-E_{GS}=\delta E^{(3)}$. For $n=3,4$, instead, $E_{n=3,4}^{(3)}-E_{GS} \approx 6 \delta E^{(3)}$ in such a way that, to provide a rough estimate of the previous expressions, we can disregard the contributions of these higher energy states. See Appendix \ref{app:Kondodet} for the explicit expressions of these operators and Appendix \ref{app:gamma} for an estimate of the truncation to $n=1,2$.

Based on the previous equations, the Kondo Hamiltonian $H_K$ acquires the following form, up to a global energy shift:
\begin{multline} \label{Kondo1}
H_{K} = \sum_\tau \left[ J^{(\perp)}_\tau l^\dag_{\tau L}l^\dag_{\tau R}S^- + {\rm{H.c.}} \right] + \sum_{\tau,\eta} V_{\tau\eta} l^\dag_{\tau \eta} l_{\tau \eta} S^z \\
+ \sum_\tau \left[J^{\rm(cot)}_\tau l^\dag_{\tau L}l_{\tau R}S^z + {\rm H.c.} \right] -  \frac{1}{2}\sum_{\tau \eta} V_{\tau\eta} S^z\,.
\end{multline}
In this expression, $J_\tau^{(\perp)}$ is the amplitude of processes in which pairs of electrons tunnel between the poor man's tetron and the leads. As the symmetry sectors $(4,-)$ and $(2,-)$ acquire energies considerably higher than $E_{GS}$, these events must preserve the $P_\Dnt$ parity of the poor man's tetron and are thus restricted to effective crossed Andreev processes in which Cooper pairs are formed by pairing electrons from the two sides of the same nanowire. In this respect, the $\eta$ degree of freedom plays the role of an effective spin in the corresponding Kondo problem and $J^{(\perp)}_\tau$ defines the related exchange term amplitude, with the nanowire $\tau$ labelling the Kondo channel. This interaction constitutes the foundation of the resulting two-channel Cooper pair charge Kondo model, as it acquires the same form of the exchange term in the charge Kondo problem emerging in spinful proximitized nanowires \cite{Pustilnik2017}. We observe, however, that in our model the quantum dots are fully polarized, such that local Andreev processes are suppressed and all the pairing interactions between electrons happen non-locally.

The other terms in Eq. \eqref{Kondo1} represent instead charge preserving processes. The $V$ interaction refers to the coupling of the lead density to the poor man's tetron charge and it results from the terms $\Xi^{\alpha \alpha} - \Upsilon^{\alpha \alpha}$ in Eq. \eqref{Kondo0}. Due to the PH symmetry we have:
\begin{equation}
V_{\tau\eta} = g_\Xi\frac{J_{\tau \eta}^2}{\delta E^{(3)}} \label{V1}  \,,
\end{equation}
where $g_\Xi$ is a numerical coefficient which results from the eigenstate wavefunctions and depends on the poor man's tetron's parameters (see Appendix \ref{app:Kondodet}).
The amplitude $J_\tau^{(\rm cot)}$ represents cotunneling events between the two sides of the same nanowire, and it acquires a sign given by the charge state of the poor man's tetron. Finally, the last term of Eq. \eqref{Kondo1} derives from the $\Upsilon^{\alpha \alpha}$ terms of Eq. \eqref{Kondo0} and corresponds to a Zeeman splitting  emerging from the suppression of the processes populating the sectors with $N_t=1,5$.

The first line of Eq. \eqref{Kondo1} is equivalent to a two-channel charge Kondo problem. This can be easily seen by a suitable charge conjugation of the right leads, analogously to the case of spinful Cooper pair charge Kondo models \cite{Garate2011}. This charge conjugation reads: 
\begin{equation} \label{conjugacy}
\tilde{l}_{\tau L}= l_{\tau L}\,, \quad \tilde{l}_{\tau R}= l^\dag_{\tau R}. 
\end{equation}
Then, we can define the effective leads spin operators:
\begin{equation}
\tilde{s}_\tau^{i} = \tilde{l}^\dag_{\tau \eta} \sigma^i_{\eta \eta'} \tilde{l}_{\tau \eta'}\, \label{Spin} ,
\end{equation}
where the $\sigma^i$ are Pauli matrices, and we label by $i=0$ the identity.
The Hamiltonian $H_K$ can thus be rewritten as:
\begin{multline} \label{Kondo2}
H_K = \sum_\tau J_\tau^{(\perp)}\left(  \tilde{s}_\tau^+S^- +\tilde{s}_\tau^-S^+ \right) + \sum_\tau J_\tau^{(z)} \tilde{s}_\tau^zS^z \\
+ \sum_\tau \left[J^{\rm(cot)}_\tau \tilde{l}^\dag_{\tau L}\tilde{l}^\dag_{\tau R}S^z + {\rm H.c.} \right] +\sum_\tau W_{\tau } \tilde{s}^0_\tau S^{z}  - h S^z\,.
\end{multline}
In this expression, $J^{(\perp)}_\tau$ and $J_\tau^{(z)}=(V_{\tau L} + V_{\tau R})/2$ are the Kondo couplings of the equivalent two-channel Kondo model and they are both positive.
The $W$ terms couple the density of the leads $\tilde{s}^0$ (after charge conjugation) with the charge degree of freedom $S^z$, and can also be interpreted as backscattering terms mediated by $S^z$. Both these terms and the Zeeman amplitude $h$ are proportional to $\sum_{\tau} \left(J_{\tau L}^2 -  J_{\tau R}^2\right)$ and they vanish when the model is symmetric under a left/right reflection. Finally, the cotunneling terms $J^{\rm (cot)}_\tau$ do not have a clear analog in the mapping into the two-channel Kondo model as, after the charge conjugation \eqref{conjugacy}, they correspond to Andreev processes.

The spatially symmetric case $J_{\tau \eta}=J$ allows us to get some preliminary insight about the two-channel Kondo fixed point emerging from Eq. \eqref{Kondo2}: $J^{(\perp)}$ and $J^{(z)}$ respectively represent the exchange and antiferromagnetic coupling between leads and impurity spin, whereas $J^{\rm (cot)}$ constitutes a marginal perturbation in the renormalization group (RG) sense. For symmetric couplings $W=h=0$, such that the Hamiltonian \eqref{Kondo2} can flow, based on a RG approach, towards strong coupling, thus revealing the non-Fermi liquid physics of the two-channel Kondo model. 

Finally we emphasize that we are interested in a regime with $t_{\rm COT}(0) \ll \Delta_{\rm CAR}$ ($E_C \ll \Delta$ and sizeable $\alpha$). These conditions yield that the cotunneling strength between the leads, $J^{\rm{(cot)}}$, is much weaker than $J^{(\perp)}$ and $J^{(z)}$ (see Appendix \ref{app:Kondodet}). Therefore, in the following, we will mostly focus on the first line of the Kondo Hamiltonian \eqref{Kondo2}, treating the other terms as perturbations.

\section{The weak coupling regime} \label{sec:weak}

\subsection{RG analysis at weak coupling}

To understand the low-temperature properties of the system, we consider first a RG analysis to reveal the scaling of the coupling constants in Eq. \eqref{Kondo2} from the weak-coupling perspective.
The Cooper pair charge Kondo Hamiltonian that we analyze is the following:
\begin{multline} \label{Kondo3}
H_K = \sum_\tau J_\tau^{(\perp)}\left(  \tilde{s}_\tau^+S^- +\tilde{s}_\tau^-S^+ \right) + \sum_\tau J_\tau^{(z)} \tilde{s}_\tau^zS^z \\
+ \sum_\tau \left[J^{\rm(cot)}_\tau \tilde{l}^\dag_{\tau L}\tilde{l}^\dag_{\tau R}S^z + {\rm H.c.} \right].
\end{multline}
The first two terms correspond to a standard 2-channel ($\tau$) Kondo model with antiferromagnetic couplings $J^{z}_\tau >0$, and represent a Cooper pair charge Kondo effect \cite{Garate2011,Papaj2019}.
The simplest RG analysis of the Kondo Hamiltonian \eqref{Kondo3} can be performed based on the standard Anderson's poor man scaling. At second order, we obtain:
\begin{align}
&\frac{ {\rm d}J_\tau^{(z)}}{{\rm d} l} = \rho \left(J^{(\perp)}_\tau\right)^2, \label{RGz}   \\ 
&\frac{ {\rm d}J_\tau^{(\perp)}}{{\rm d} l} = 4\rho J_\tau^{(z)} J^{(\perp)}_\tau\,, \label{RGperp} \\
&\frac{ {\rm d}J_\tau^{\rm (cot)}}{{\rm d} l}=0\,, \label{RGcot}
\end{align}
where $l$ is the renormalization flow parameter and $\rho$ labels the density of states of the leads at the Fermi level (see Appendix \ref{app:RG} for more detail). The first two equations describe the standard 2-channel Kondo RG flow \cite{Pustilnik2004}, and they match the renormalization equations of the Cooper pair charge Kondo effect in superconducting grains \cite{Garate2011}. The additional cotunneling terms at second order do not renormalize the other couplings and are marginal, such that they do not affect the flow from weak to strong coupling.

\begin{figure}[t]
\includegraphics[width=\columnwidth]{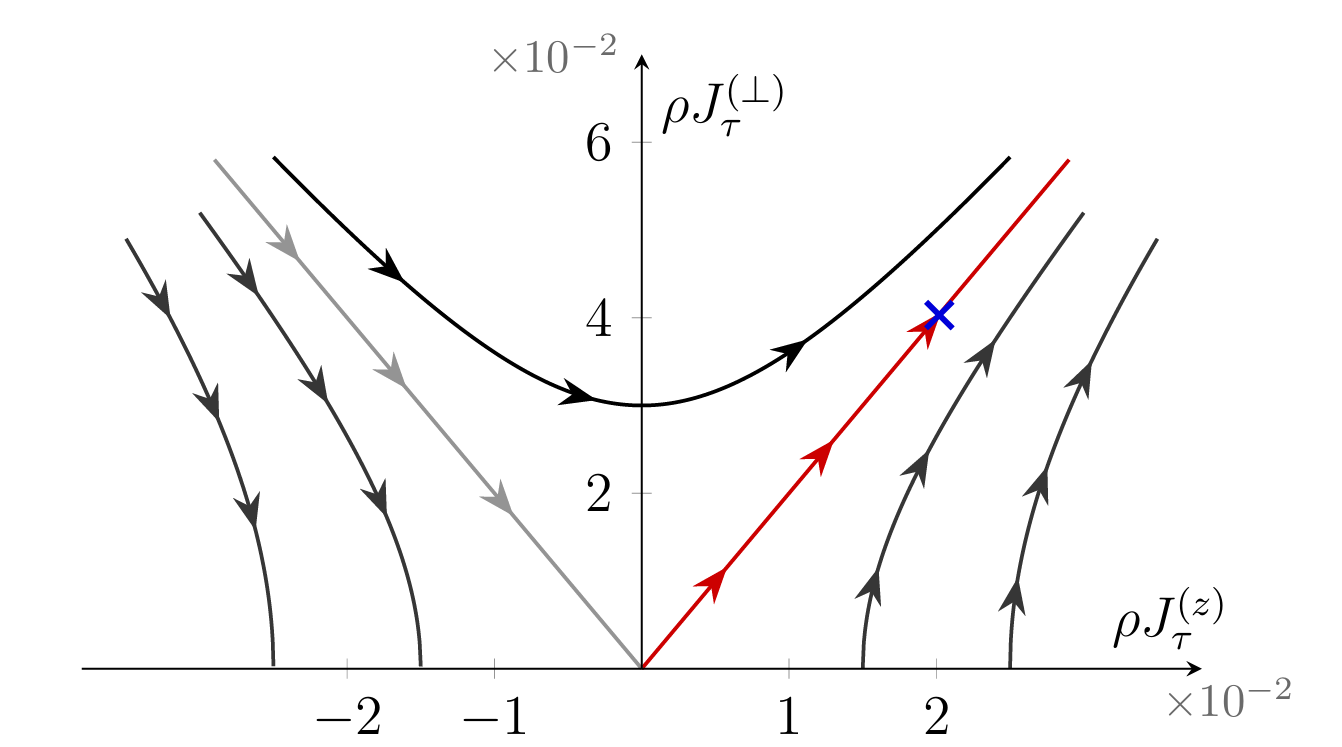}
\caption{RG flow of the Kondo couplings based on the second-order equations. A typical value of the bare couplings $\rho J^{(\perp)}_{\tau}\simeq0.04, \; \rho J^{(z)}_{\tau}\simeq0.02$ is indicated in the plot by a blue cross. It corresponds to the estimates in Eq. (\ref{Perp_0},\ref{Zeta_0}) in a strong coupling regime, $J\simeq \delta E^{(3)}$, and for the following parameter choice: $\Delta=2.5t, \; E_C=t, \; \alpha=\pi/6, \rho=3.4$meV$^{-1}$.} \label{fig:RGFlow}
\end{figure}
These renormalization equations describe the flow of the coupling constants for systems with an emerging mirror symmetry $\eta \leftrightarrow -\eta$, such that $W$ and $h$ in Eq. \eqref{Kondo2} are suppressed. The resulting RG flow for an arbitrary channel $\tau$ is depicted in Fig. \ref{fig:RGFlow}. As in the standard Kondo model \cite{Hewson_1993}, the second-order equations (\ref{RGz},\ref{RGperp}) preserve the invariants $\mathcal{I}_\tau=\left(J_{\tau}^{(\perp)}\right)^2- 4\left(J_{\tau}^{(z)}\right)^2 $ which determine the hyperbolic pattern of the flow. In our model, the bare coupling constants correspond to the antiferromagnetic regime $J^{(z)}_\tau>0$, such that the flow proceeds towards strong coupling. The difference $J^{(\perp)}_\Upt - J^{(\perp)}_\Dnt$ between the couplings of the two nanowires, however, is relevant, as in the standard two-channel Kondo model \cite{Pustilnik2004}. This implies that any breaking of the nanowire symmetry $\tau \leftrightarrow -\tau$ in Eq. \eqref{Kondo3} yields, at sufficiently low energies, a ground state with the charge degree of freedom $N_t$ coupled with the leads of the dominant nanowire only, whereas the leads connected to the other nanowire become decoupled. This situation corresponds to a single-channel Kondo fixed point. The physics of the two-channel Kondo model, and the related non-Fermi liquid behavior, emerges at the critical point that separates the phases with dominant $\tau=\Upt$ and $\tau=\Dnt$ channels. If the nanowire anisotropy is low, however, the dynamics of the system is expected to be governed by the two-channel Kondo behavior for sizeable ranges of temperature and external bias.

At the critical point invariant under the $\tau \leftrightarrow -\tau$ symmetry, the RG equations (\ref{RGz},\ref{RGperp}) determine the onset of a unique two-channel Kondo temperature for the whole system. For the case $J^{(\perp)}> 2 J^{(z)}$ we obtain:
\begin{equation} \label{TK}
T_{K}\approx \delta E^{(3)}\ee^{-\frac{1}{2\rho \sqrt{\mathcal{I}}}\arccos \frac{2J^{(z)}(0)}{J^{(\perp)}(0)}}\,,
\end{equation}
with $\mathcal{I}=\mathcal{I}_\tau$ at the symmetric critical point.
In Appendix \ref{app:RG} we derive this estimate and we provide the corresponding expression for the case $J^{(\perp)}< 2 J^{(z)}$. In the regime $E_C \ll \Delta$, we expect $2J^{(z)} \approx J^{(\perp)}$ (see Appendix \ref{app:Kondodet}) such that both the arccosine and $\sqrt{I}$ vanish in Eq. \eqref{TK} and the Kondo temperature can be approximated by 
\begin{equation} \label{TK2}
T_K \approx \delta E^{(3)} \ee^{-\frac{1}{2\rho J^{(\perp)}}}.
\end{equation}

Besides the breaking of the nanowire  symmetry $\tau \leftrightarrow -\tau$, also the breaking  of the left/right mirror symmetry $\eta \leftrightarrow -\eta$ drives the system away from the two-channel Kondo fixed point. The effect of the left/right lead anisotropy is actually stronger as it is translated into the introduction of the Zeeman term $h$ in Eq. \eqref{Kondo2}, which is relevant in the RG sense.

In particular, the left/right anisotropy yields both the Zeeman and the $W$ perturbations in Eq. \eqref{Kondo2}, which scale as:
\begin{align}
&\frac{ {\rm d}h}{{\rm d} l}= h\,, \label{RGh}\\
&\frac{ {\rm d}W_\tau}{{\rm d} l} = 0\,, \label{RGW}
\end{align}
where the poor man's scaling of $W$ is obtained in a way similar to $J^{(z)}$ (see Appendix \ref{app:RG}). Given its first-order renormalization, the main limitation to the onset of the non-Fermi liquid two-channel charge Kondo regime is constituted by the Zeeman perturbation $h$. Its main effects, however, must be evaluated by considering the strong coupling regime (see Sec. \ref{sec:strong}).

\subsection{Andreev conductance at weak coupling}

We now investigate the implications of the onset of the Cooper pair charge Kondo effect on the transport properties of the system in a weak-coupling regime. The transport features of mesoscopic superconducting islands coupled to normal state leads have been extensively studied \cite{Garate2011,Pustilnik2017,Papaj2019} and our analysis follows the steps adopted in Ref. \cite{Garate2011}.

Charge transport in the poor man's tetron is mediated by two distinct processes, namely CAR and cotunneling. In the following, we will focus in particular on the electronic transport mediated by CAR processes which allow for the net transfer of Cooper pairs between the two nanowires, as opposed to local Andreev processes, which are instead suppressed by the polarization of the quantum dots in our setup. 
Furthermore, cotunneling processes, mediated by $J^{(\text{cot})}$, do not contribute to the interwire transport, and we will therefore neglect them in the following discussion.

\begin{figure}[t]
    \centering
    \includegraphics[width=\columnwidth]{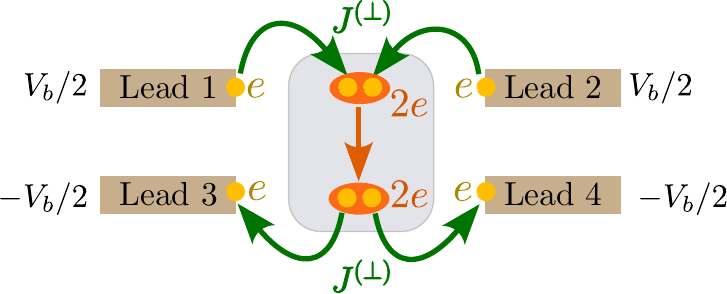}
    \caption{Sketch of the processes generating the interwire Andreev conductance. The two nanowires are subject to a voltage bias $V_b$ and the Andreev conductance $G_{\rm CAR}$ accounts for processes in which a Cooper pair is formed by electrons in leads 1 and 2, and split into two correlated electrons injected in leads 3 and 4.}
    \label{fig:GCAR}
\end{figure}

The transport of pairs of electrons between the two wires is indeed generated by second-order processes of the $J^{(\perp)}$ terms of the Hamiltonian \eqref{Kondo3}, which dominate the dynamics of the system in the two-channel strong coupling regime (see Fig. \ref{fig:GCAR}). In such processes, a Cooper pair is formed from two electrons originating from the two leads connected to the first nanowire, and it is then split into two outgoing electrons in the other two leads. These processes produce correlated pairs of outgoing electrons and yield a net interwire current when introducing a voltage bias $V_b$ between the two. This corresponds to a voltage bias between the two charge Kondo channels and it acquires the following form:
\begin{equation}
    H_{\rm{bias}}=\frac{e V_{b}}{2}\sum_{\eta=\rm{L,R}}\Psi^{\dagger}_{\eta}\tau_{z}\Psi_{\eta}\ \label{bias},
\end{equation}
where $\Psi_{\eta}=(l_{\rm{u},\eta},l_{\rm{d},\eta})^{\rm{T}}$ is a two-component spinor and $\tau_{z}$ is the Pauli matrix acting on the nanowire degree of freedom. We now define the Andreev current $I^{\rm{A}}$ as the average between the expectation values of the ingoing current from the top nanowire and the outgoing current from the bottom one:
\begin{equation} \label{AndreevC0}
I^{\rm{A}}=\frac{1}{2}\sum_{\tau}I_{\tau}^{\rm{A}}\ ,
\end{equation}
where we defined $I_{\tau}^{\rm{A}}=\langle \hat{I}^{\text{A}}_{\tau}\rangle=2ie\tau\langle[H_K, N_{\tau\eta}]\rangle$ with $N_{\tau\eta}$ the number operator on lead $(\tau,\eta)$.  The commutator must be evaluated considering only the first term in Eq. \eqref{Kondo3}, which is responsible for CAR processes. We obtain:
\begin{equation}
    I_{\tau}^{\rm{A}}=2ie\tau J_{\tau}^{(\perp)}\left[\langle l_{\tau R}l_{\tau L}S^{+}\rangle-\langle l^{\dagger}_{\tau L}l^{\dagger}_{\tau R}S^{-}\rangle\right]\ \label{AndreevC} , 
\end{equation} 
with $\tau=\pm1$ for the top and bottom nanowire respectively. The Andreev current between the two nanowires can equivalently be reformulated as a spin current in the two-channel Kondo analogy, after applying the PH transformation in Eq. (\ref{conjugacy}). In particular, the voltage bias in Eq. (\ref{bias}) now plays the role of a Zeeman term for the two channels acting on the $\eta=\rm{L,R}$ degree of freedom. Indeed, we find:
\begin{equation}
    H_{\rm{bias}}=\frac{eV_{b}}{2}\left(\tilde{s}^{z}_{u}-\tilde{s}^{z}_{d}\right)\ ,
\end{equation}
where $\tilde{s}^{z}_{\tau}$ denotes the lead spin operator in Eq. (\ref{Spin}). Analogously, the Andreev current in Eq. (\ref{AndreevC}) is mapped into a spin current $\tilde{I}^{\rm{S}}$ given by:
\begin{equation}
\tilde{I}^{\rm{S}}_{\tau}=2ie\tau J_{\tau}^{(\perp)}\left[\langle\tilde{s}^{-}_{\tau}S^{+}\rangle-\langle\tilde{s}^{+}_{\tau}S^{-}\rangle\right].    
\end{equation}

To determine the Andreev conductance at weak coupling, we adopt a linear response analysis, following the discussion presented in Ref. \cite{Garate2011}. In the regime of linear response, we can define the Andreev conductance as follows:
\begin{equation}
 I^{\text{A}}=G_{\text{CAR}} V_b \ ,
\end{equation}
where $I^{\text{A}}$ is the (averaged) Andreev current defined in Eq. \eqref{AndreevC0}, and $V_b$ the voltage bias.

The Kubo formalism \cite{Mahan} allows us to extract the zero-bias Andreev conductance from the current-current correlation function in imaginary time:
\begin{equation}
    G_{\text{CAR}}=\lim_{\omega\to0}\frac{1}{i\omega}\int^{\beta}_{0}\rm{d}\tau\  \ee^{i\omega_n\tau}\langle T_{\tau}\hat{I}^{\text{A}}(\tau)\hat{I}^{\text{A}}(0)\rangle \label{Kubo} \ ,
\end{equation}
where $\rm{T}_{\tau}$ denotes time ordering in the imaginary time $\tau$ and the correlator in the Kubo formula is consistent with a low voltage bias perturbation ($V_b \ll \delta E^{(3)}$) within the second-order perturbative approximation adopted to derive Eq. \eqref{Kondo3}. The integral can be computed using standard techniques (see Ref. \cite{Mahan}) by employing the expression of the Andreev current operator in Eq. (\ref{AndreevC0}). We find:
\begin{equation} \label{CARG}
G_{\text{CAR}}=2\tilde{G}_{0}\int d\epsilon \left(-\partial_{\epsilon} n_{F}\right)\rho^{2}\left(J^{(\perp)}\right)^{2},
\end{equation}
where $n_{F}$ and $\rho$ denote respectively the Fermi-Dirac distribution function and the density of states in the leads, and we defined $\tilde{G}_{0}\equiv 2\pi^2e^2/h$. Based on the solutions of the RG equations in the weak coupling limit,  we expect the Kondo zero-bias peak to decay with temperature as
\begin{equation}\label{GCART}
G_{\rm CAR} \simeq \tilde{G}_0 \frac{1}{2\ln^2\left(T/T_{K}\right)}\,,
\end{equation}
in a suitable scaling regime $T_K < T < \delta E^{(3)}$ (see Appendix \ref{app:RG}). This temperature dependence matches the standard charge Kondo behavior obtained with a rate equation approach by considering sequential tunneling processes with the renormalized rate provided by the symmetric couplings $J^{(\perp)}$ at the energy scale $T$ \cite{Matveev1995}.

\section{Non-Fermi liquid signatures at strong coupling} \label{sec:strong}

\subsection{Emery--Kivelson solution}

We now address the strong coupling regime by adopting a bosonized description of $H_K$ in Eq. \eqref{Kondo1} and following the Emery--Kivelson construction for two-channel Kondo systems \cite{EmeryKivelson}. For simplicity, we begin by considering the channel-symmetric case $J^{(\lambda)}_{\Upt}=J^{(\lambda)}_{\Dnt}\equiv J^{(\lambda)}$ with $\lambda=\perp, z$, and assuming non-interacting leads. 

Close to the Fermi energy, the four leads are described by incoming and outgoing chiral fermions on the half-line,
\begin{equation}
\begin{aligned}
H_{\rm leads}=i v_F\sum_{\tau,\eta}\int_0^{+\infty} dx\,
&\left(
\psi^\dagger_{\tau\eta,{\rm in}}
\partial_x\psi_{\tau\eta,{\rm in}}\right.\\
&\left.-\psi^\dagger_{\tau\eta,{\rm out}}\partial_x\psi_{\tau\eta,{\rm out}}\right),
\label{EK-leads}
\end{aligned}
\end{equation}
where we assumed to have the same Fermi velocity $v_F$ for all the external leads.
We bosonize the chiral fields according to \begin{equation}
\psi_{\tau\eta,{\rm in/out}}(x)
=
\frac{\kappa_{\tau\eta}}{\sqrt{2\pi a}}\,
\ee^{i[\phi_{\tau\eta}(x)\pm\theta_{\tau\eta}(x)]},
\label{EK-bos}
\end{equation}
where $a$ is a short-distance cutoff and $\kappa_{\tau\eta}$ labels suitable
Klein factors. The canonical bosonic fields obey
\begin{equation}
\left[
\phi_{\tau\eta}(x),\theta_{\tau'\eta'}(y)
\right]
=
i\pi\Theta(y-x)\delta_{\tau\tau'}\delta_{\eta\eta'},
\end{equation}
where $\Theta$ is the Heaviside step function and the operator $-\frac{\partial_x \theta_{\tau \eta}}{\pi}$ indicates the fluctuations of the charge density in the related lead.
The four leads are then described by four independent non-interacting Luttinger liquids on the half-line. It is convenient to introduce their linear combinations \cite{EmeryKivelson} 
\begin{align} 
\phi_c&=\frac{1}{2}\sum_{\tau,\eta}\phi_{\tau\eta}, & \phi_s&=\frac{1}{2}\sum_{\tau,\eta}\eta\phi_{\tau\eta}, \nonumber\\ \phi_t&=\frac{1}{2}\sum_{\tau,\eta}\tau\phi_{\tau\eta}, & \phi_{st}&=\frac{1}{2}\sum_{\tau,\eta}\tau\eta\phi_{\tau\eta}, 
\label{EK-fields} 
\end{align} 
and analogously for the $\theta_a$ fields.

The impurity couples only to the boundary operators of the lead fields which we represent as $l_{\tau\eta}=\sqrt{a}\psi_{\tau \eta}(0)$. The matching between incoming and outgoing fields is
not fixed at this stage, but is determined dynamically by the
impurity interaction. The bosonization of the impurity interaction can now be carried out directly in terms of the bosonic fields in Eq. \eqref{EK-fields}. Based on Eq. \eqref{EK-bos}, both the crossed Andreev terms $J^{(\perp)}$ and the cotunneling interaction $J^{(\rm cot)}$ associated with the nanowire $\tau$ acquire a dependence on the vertex operators $\exp\left[\pm i\left(\theta_{\tau L}+\theta_{\tau R}\right)\right]$. These operators correspond to the fermionic parities $P_{\tau}$ and are therefore fixed in the $P_{\Upt}=P_{\Dnt}=+1$ symmetry sector considered here. Similarly, the Klein factors enter the Hamiltonian only through the bilinears $i\kappa_{\tau L}\kappa_{\tau R}$.  These bilinears commute with the Kondo Hamiltonian and can consequently be replaced by their eigenvalues. Hence, the bosonized impurity Hamiltonian becomes
\begin{multline}
H_{\rm imp}=\frac{J^{(\perp)}}{\pi}\cos \phi_t(0)\left[\sin \phi_c(0) S^x+\cos \phi_c(0) S^y\right]\\
-\frac{2J^{(z)}}{\pi}a\partial_x\theta_c(0)S^z+\frac{2J^{\rm(cot)}}{\pi}\cos\phi_{st}(0)\sin\phi_s(0)S^z.
\label{EK-bosonized-H}
\end{multline} 
This Hamiltonian corresponds to the standard Emery-Kivelson description of the two-channel Kondo effect via the particle-hole transformation in Eq. \eqref{conjugacy}. The first two terms constitute the Cooper pair charge Kondo Hamiltonian. The impurity charge flip couples indeed to the total charge field $\phi_c$, whereas $\phi_t$ distinguishes the two nanowires and therefore carries the Kondo channel degree of freedom. The last term originates from the charge-dependent cotunneling processes. It is parametrically weak in the regime $J^{\rm(cot)}\ll J^{(\perp)},J^{(z)}$ considered in this work.

The dependence of the transverse exchange on the charge field can be removed by the rotation \cite{EmeryKivelson}
\begin{equation}
U=\exp\left[\frac{i}{2}\phi_c(0)S^z\right].
\label{EK-rotation}
\end{equation}
With the Pauli-matrix convention $[S^z,S^\pm]=\pm2S^\pm$, the transformed Hamiltonian $H'=U^\dagger H U$ takes the form
\begin{multline}
H'=H_{\rm leads}+\frac{J^{(\perp)}}{\pi}\cos\phi_t(0)S^y-\lambda_z\,\partial_x\theta_c(0)S^z
\\
+\frac{2J^{\rm(cot)}}{\pi}\cos\phi_{st}(0)\sin\phi_s(0)S^z ,
\label{EK-rotated-H}
\end{multline}
up to an additive constant. The coupling
\begin{equation}
\lambda_z\equiv\frac{2J^{(z)}a}{\pi}-\frac{v_F}{2}
\end{equation}
measures the residual coupling between the impurity and the charge density. 

As shown in Eqs. \eqref{RGz} and \eqref{RGperp}, and depicted in Fig.~\ref{fig:RGFlow}, at the decoupled weak-coupling fixed point, the couplings $J^{(\perp)}$ and $J^{(z)}$ are marginal at the tree level and jointly generate the perturbative RG flow toward the two-channel Kondo fixed point. Therefore, to isolate the universal strong-coupling structure, we first neglect $J^{\rm(cot)}$ and tune the longitudinal coupling to the Emery--Kivelson (EK) line, $J^{(z)}=J^{(z)}_{*}\equiv\frac{\pi v_F}{4a}$. At this point the charge sector decouples from the impurity, while the transverse coupling acts only on the $t$ sector. The latter can be refermionized and unfolded into a single chiral fermion on the full line
\begin{equation}
\psi_t(x)=\begin{cases}
    \psi_{t,\rm{out}}(x)\;,\quad x>0\\
    \psi_{t,\rm{in}}(-x) \;,\quad x<0
\end{cases}
\label{EK-refermionization}
\end{equation}
where $\psi_{t,\rm{in/out}}=\frac{\kappa_t}{\sqrt{2\pi a}}e^{i(\phi_t(x)\pm\theta_t(x))}$.
By following the construction in Ref. \cite{EmeryKivelson}, we represent the impurity by two Majorana operators $\gamma_1$ and $\gamma_2$ such that $S^z=i\gamma_1\gamma_2$, $S^y=i\kappa_t\gamma_1$ and $S^x=i\kappa_t\gamma_2$. The resulting EK Hamiltonian is therefore quadratic,
\begin{multline}
H_{*}=H_0^{c,s,st}-i v_F\int_{-\infty}^{+\infty}dx\,\psi_t^\dagger(x)\partial_x\psi_t(x)
\\
+i J^{(\perp)} \sqrt{\frac{a}{2\pi}}\left[\psi_t(0)+\psi_t^\dagger(0)\right]\gamma_1 ,
\label{EK-quadratic-H}
\end{multline}
where we exploited the relation $\cos(\phi_t) \propto \psi_t^\dagger(0)+\psi_t(0)$, based on the parity constraints $P_{\rm u}=P_{\rm d}=1$.
Only the Majorana $\gamma_1$ hybridizes with the lead continuum, with a characteristic width $\Gamma= (J^{(\perp)})^2a/(\pi v_F)$, whereas $\gamma_2$ remains decoupled. The latter gives the residual impurity entropy $S_{\rm imp}=\frac{1}{2}\ln 2$ characteristic of the two-channel Kondo fixed point.

The boundary scaling properties of the operators at the Emery--Kivelson fixed point can be extracted by integrating out the gapless $t$-sector fermion \cite{Altland2023}. This produces a zero-dimensional effective Euclidean action for the decoupled Majorana combination $\eta_t(0) = i\left(\psi_t^\dagger(0)-\psi_t(0)\right)$ and the two impurity Majoranas \cite{EmeryKivelson, Sengupta1994}, 
\begin{multline}
S_{*}=\frac{1}{4\beta}\sum_{\omega_n}\gamma_{1,-\omega_n}\left[-i\omega_n-i\Gamma\,{\rm sgn}(\omega_n)\right]\gamma_{1,\omega_n}\\
+\frac{1}{4\beta}\sum_{\omega_n}\gamma_{2,-\omega_n}\left(-i\omega_n\right)\gamma_{2,\omega_n}\\
+\frac{1}{2\beta}\sum_{\omega_n}\eta_{t,-\omega_n} \left[i v_F {\rm sgn}(\omega_n)\right] \eta_{t,\omega_n},
\label{EK-0D-action}
\end{multline}
where $\omega_n=(2n+1)\pi T$ are fermionic Matsubara frequencies. The derivation of this boundary action, obtained by integrating out the gapless bulk modes, is reported in Appendix~\ref{app:SC-conductance}.
The first Majorana acquires a dissipative self-energy from its hybridization with the continuum, whereas $\gamma_2$ remains free.
At frequencies $|\omega_n|\ll\Gamma$, the propagator of the hybridized Majorana approaches $-i\Gamma^{-1}{\rm sgn}(\omega_n)$. 
Its long-time Euclidean correlation function consequently behaves as 
\begin{equation}
G_{\gamma_1}(\tau)=\left\langle T_{\tau}\gamma_1(\tau)\gamma_1(0) \right\rangle_{*} \simeq \frac{1}{\Gamma} \frac{2}{\beta\sin(\pi \tau/\beta)},
\label{EK-gamma-correlation}
\end{equation}
for $\Gamma^{-1}\ll\tau<\beta$, showing that $\gamma_1$ has boundary scaling dimension $1/2$ at the EK fixed point. The correlation function of the decoupled Majorana $\gamma_2$, instead, does not decay at long times corresponding to vanishing boundary scaling dimension. Consequently, the impurity operator $S^z=i\gamma_1\gamma_2$ inherits the long-time scaling of the hybridized Majorana and therefore acquires a boundary scaling dimension $1/2$.

We may now consider the residual longitudinal interaction $H_z=- \lambda_z \partial_x\theta_c(0)S^z$ as a perturbation to the exactly solvable quadratic EK Hamiltonian. At the EK fixed point, the charge sector decouples from the impurity and remains free, so that the boundary density $\partial_x\theta_c$ has boundary scaling dimension $1$. Overall, the boundary operator entering in $H_z$ is thus irrelevant at the EK fixed point, with boundary scaling dimension $3/2$ and RG eigenvalue $1-\Delta_{*}[H_z]=-1/2$. Nevertheless, it is the least irrelevant operator and therefore it determines the leading deviation from the fixed-point transport properties discussed in the next  \cite{Affleck1991}.

\subsection{Andreev conductance at strong coupling}
\label{sec:EK-conductance}
We now make use of the EK description of our model to determine the universal Andreev conductance at the two-channel Kondo fixed point and its leading low-temperature correction. The Andreev current introduced in Eq. \eqref{AndreevC0} can be expressed in terms of the $t$-sector fermion and the hybridized Majorana. After the rotation \eqref{EK-rotation}, we obtain
\begin{equation}
I^{\rm A}=2eJ^{(\perp)}\sqrt{\frac{a}{2\pi}}\left[\frac{\psi_t^\dagger(0)-\psi_t(0)}{2}\right]\gamma_1.
\label{EK-current-refermionized}
\end{equation}
Equation~\eqref{EK-current-refermionized} shows that the Andreev current probes directly the Majorana resonance associated with the two-channel Kondo fixed point.

On the EK line, the Hamiltonian is quadratic and the current correlation function entering Eq. \eqref{Kubo} can be evaluated employing Wick contractions. In particular, Eq.~\eqref{EK-current-refermionized} involves the free $t$-sector fermion and the dressed propagator of $\gamma_1$. To recover the generic low-temperature behavior of the two-channel Kondo fixed point, we restore the residual longitudinal interaction and refermionize the decoupled charge sector according to $:\!\psi_c^\dagger\psi_c\!:=-\frac{1}{\pi}\partial_x\theta_c $. The perturbation of the EK action $S_*$ is then
\begin{equation}
\delta S=\pi\lambda_z\int_0^\beta d\tau\,\mathcal O_{3/2}(\tau),
\qquad\mathcal O_{3/2}=i:\!\psi_c^\dagger\psi_c\!:\gamma_1\gamma_2 .
\label{EK-irrelevant-operator}
\end{equation}
The Euclidean current correlation can therefore be expanded around the EK fixed point as
\begin{multline}
\left\langle T_\tau I^{\rm A}(\tau)I^{\rm A}(0)\right\rangle=\left\langle T_\tau I^{\rm A}(\tau)I^{\rm A}(0)\right\rangle_*\\
+\frac{\pi^2\lambda_z^2}{2}\int_0^\beta d\tau_1\,d\tau_2\,\left\langle T_\tau I^{\rm A}(\tau)I^{\rm A}(0) \mathcal O_{3/2}(\tau_1)\mathcal O_{3/2}(\tau_2)\right\rangle_{*,c}\\
+O(\lambda_z^4).
\label{EK-current-expansion}
\end{multline}
where the odd-order corrections in $\lambda_z$ vanish because the decoupled charge density has zero expectation value at the fixed point.

The leading order current correlation function factorizes as product of the $\eta_t$ and $\gamma_1$ free propagators 
\begin{equation}
\left\langle T_\tau I^{\rm A}(\tau)I^{\rm A}(0)\right\rangle_*=e^2 \frac{a}{2\pi}(J^{(\perp)})^2G_{\eta_t}(\tau)G_{\gamma_1}(\tau).
\label{EK-fixed-current-correlation}
\end{equation}
Given the long times correlations and after the analytic continuation entering the Kubo formula, one obtains the universal fixed-point value
\begin{equation}
G_{\rm CAR}(T=0)=G_*=\frac{2e^2}{h}.
\label{EK-fixed-conductance}
\end{equation}
The corresponding frequency-space Kubo calculation is presented in Appendix~\ref{app:SC-conductance}. Analogously to the other Cooper pair charge Kondo effects \cite{Garate2011,Pustilnik2017}, we observe that such universal value corresponds to a fractional conductance for Cooper pairs as $G_* = \frac{1}{2}\frac{(2e)^2}{h}$.

\subsection{Two-channel Kondo perturbations}
At finite temperature, the quadratic EK theory yields only analytic corrections to the fixed-point result, beginning at order $T^2$. However, at a generic point away from the exactly solvable line, the dominant correction is generated by the dimension-$3/2$ operator in Eq.~\eqref{EK-irrelevant-operator}.
Evaluating the second-order term in Eq.~\eqref{EK-current-expansion}, we obtain, for $T\ll\Gamma$,
\begin{equation}
G_{\rm CAR}(T)=G_*\left[1-\frac{\pi^3}{8\, v_F^2}\lambda_z^2\left(\frac{T}{\Gamma}\right)+ O\left(\frac{T^2}{\Gamma^2}\right)\right].
\label{EK-linear-conductance}
\end{equation}
The numerical coefficient in Eq.~\eqref{EK-linear-conductance} is obtained in Appendix \ref{app:SC-conductance} by evaluating the connected correlator in Eq. \eqref{EK-current-expansion} with $\mathcal O_{3/2}$ perturbation insertions.

The parameter $v_F^2\Gamma/\lambda_z^2$ defines the characteristic energy scale associated with the leading irrelevant perturbation close to the EK line. By matching the coefficient of the impurity susceptibility to the corresponding Bethe--Ansatz result, Sengupta and Georges~\cite{Sengupta1994} identified this scale, up to a convention-dependent numerical factor, with the thermodynamic Kondo temperature $T_K$. Therefore, the second term in Eq.~\eqref{EK-linear-conductance} displays the characteristic form of a linear correction in $T/T_K$.

This linear suppression is the characteristic non-Fermi-liquid correction of the two-channel Kondo fixed point and dictates the low-temperature behavior of the Andreev conductance, while its high-temperature behavior is captured by the weak-coupling logarithmic suppression in Eq. \eqref{GCART}. Additionally, at low temperature, we expect a further linear suppression with the voltage bias between the nanowires.

The most relevant perturbation of the two-channel Kondo fixed point, is instead represented by the Zeeman field $hS_z$ on the impurity in Eq. \eqref{Kondo2}. At the quadratic fixed point \eqref{EK-0D-action}, the spin operator has scaling dimension $\Delta_{S_z}=1/2$ inherited by the hybridized impurity Majorana $\gamma_1$. Therefore the boundary coupling $h$ has RG scaling $y_h=1-\Delta_{S_z}=1/2$. The leading RG running at the fixed point
\begin{equation}
    \dfrac{dh}{d\ell}=\dfrac{1}{2}h, \qquad \ell =\ln(T_K/T) 
\end{equation}
sets a crossover scale $T_h$ below which the system is driven away from the two-channel Kondo non-Fermi-liquid fixed point. The solution $h(\ell) = h\, \ee^{\ell/2}$ becomes indeed of the order of the fixed point UV cut-off $T_K$ at the scale 
\begin{equation} \label{Th}
T_h = T_K \left(\frac{h}{T_K} \right)^2\,.
\end{equation}
As we discuss in the next section, this sets the most pressing experimental constraints on the observation of the non-Fermi liquid features.

\section{Perturbations and experimental considerations} \label{sec:experiments}

\begin{figure*}[t]
\includegraphics[width=\textwidth]{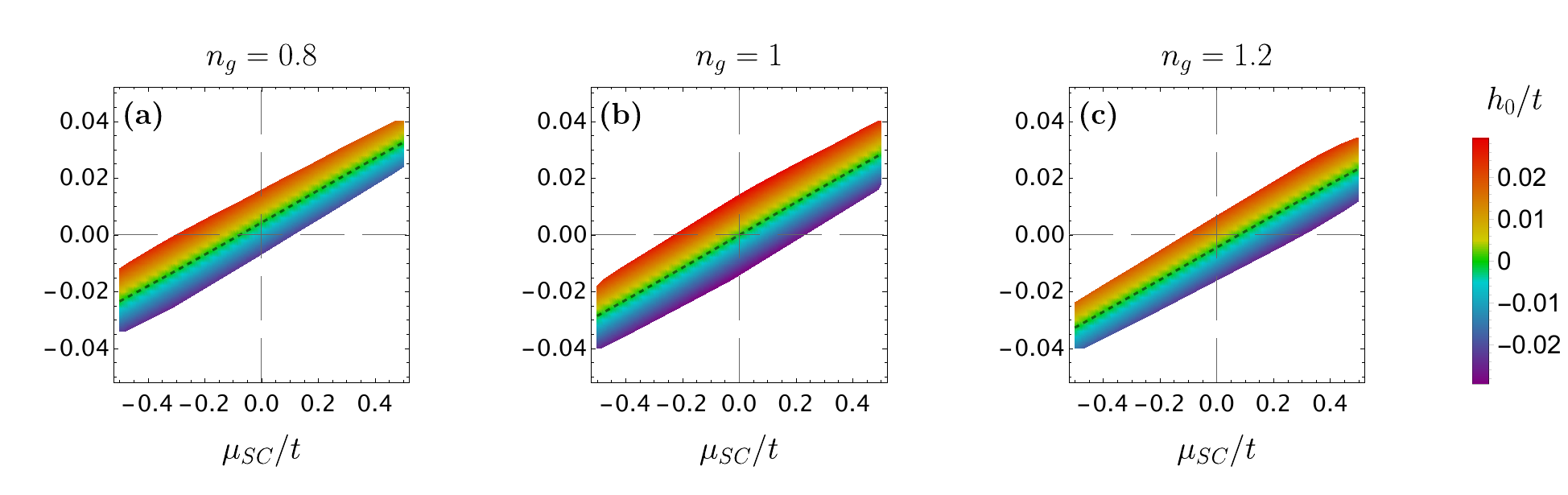}
\caption{Ground state energy splitting $h_0/t\equiv (E_{\text{GS}}^{(4,+)}- E_{\text{GS}}^{(2,+)})/t$ as a function of $\mu_{D}$ and $\mu_{\text{SC}}$ for three different values of the induced charge $n_g$, namely (a) $n_g=0.8$, (b) $n_g=1.0$ and (c) $n_g=1.2$, and for the following parameter choice: $\Delta=4t, \; E_C=0.5t, \; \alpha=\pi/6$. The point $\mu_D=\mu_{\text{SC}}=0, \ n_g=1$ corresponds to the PH symmetric point and the dashed line represents the degeneracy line where $h_0$ vanishes. Only the region where the two lowest energy states $(2,+), (4,+)$ are separated from the $(3,\pm)$ sectors by a gap $\delta E^{(3)}>3 h_0$ is shown. The results have been obtained by exact diagonalization of the full Hamiltonian $H_{\text{PMT}}$.} \label{fig:GSSplitting}
\end{figure*}

In order to provide estimates for the experimental parameter ranges required for the observation of the Cooper pair charge Kondo effect, it is first useful to observe that the Kondo model construction relies on the assumption that only the two lowest energy states $\ket{2}$ and $\ket{4}$ are populated by the dynamics.
This sets the constraint $J < \delta E^{(3)}$ which establishes the maximum lead - dot hybridization. 

The main energy scale setting the appearance of the charge Kondo effect is given by the Kondo temperature. Its estimate in Eq. \eqref{TK} relies not only on the previous perturbation theory assumption $J < \delta E^{(3)}$, but also on the weak-coupling assumption $\rho J^{(\perp)} \ll 1$. The latter condition, however, is relaxed when considering the strong-coupling regime described in Sec. \ref{sec:strong}. 

The RG considerations about the weak-coupling regime, however, are useful to determine the general behavior of $T_K$. Close to the PH symmetric point and in the perturbative tunneling regime with $t \ll \Delta - E_C$, the eigenstates of the effective Hamiltonian \eqref{hameff} are such that the ratio $R\equiv \frac{J^{(\perp)}}{2J^{(z)}}$ is very close to 1 in a broad range of the poor man's tetron parameters for isotropic couplings to the external leads $J_\tau =J$ (see Appendix \ref{app:gamma}). In this situation, $T_K$ acquires the simpler form \eqref{TK2} and depends primarily on the parameter $\rho J^{(\perp)}$.

In the weak-coupling regime $\rho J^{(\perp)} \ll 1$, the Kondo temperature in Eq. \eqref{TK2} is strongly suppressed such that $T_K \ll \delta E^{(3)}$. Therefore, to favor the critical point observation, it is necessary to tune the device away from weak coupling towards strong coupling, with $\rho J^{(\perp)} \sim 1$. In this regime, Eq. \eqref{TK2} is no longer reliable and $T_K$ increases instead towards $T_K \sim \delta E^{(3)}$ (see, for instance, the two-channel Kondo numerical RG data in Ref. \cite{Mitchell2016}). 

To reach the strong-coupling regime, therefore, we need to increase the tunneling amplitudes $J$ towards the maximal limit $\delta E^{(3)}$ set by the perturbative Kondo construction. By considering $J \sim \delta E^{(3)}$, the parameter $\rho J^{(\perp)}$ can be recast in the form $\rho J^{(\perp)} \approx 0.23 \Gamma_e / \delta E^{(3)}$ as a function of the electron tunneling rate $\Gamma_e = 2\pi \rho J^2$ between the leads and the quantum dots and the gap $\delta E^{(3)}$ (see Appendix \ref{app:Kondodet}), both of which are experimentally accessible in the experiments. Therefore, we conclude that the optimal regime for observing the two-channel Kondo dynamics is obtained when $\Gamma_e \approx 4.3 \delta E^{(3)}$.

In this regime the Kondo temperature grows towards its maximal value $T_K \sim \delta E^{(3)} \approx 0.4 \Delta_{\rm CAR}$, where the last approximation is derived from Eq. \eqref{dE3}. The experiments of the last years on hybrid quantum dots in contact with grounded aluminum superconductors demonstrated the possibility of reaching CAR amplitudes up to values of the order $\Delta_{\rm CAR}(E_C=0) \sim 40$\textmu eV \cite{Zatelli2023}. Therefore, a rough estimate of $T_K$ for the charge Kondo effect at strong coupling in the poor man's tetron would be $T_K \sim \delta E^{(3)} \approx 180 $mK. Furthermore, for a floating SC island, the effect of the typical charging energy in double nanowire devices, $E_C \sim 50$ \textmu eV \cite{Vekris2022}, enhances $\Delta_{\rm CAR}$ [see Eq. \eqref{deltacar}], such that we expect that even larger gaps $\delta E^{(3)}$ can be achieved. To obtain the strong coupling regime, then, it is necessary to control the lead - quantum dot contacts in such a way that the electron tunneling rate matches $\Gamma_e \approx 65$ \textmu eV.

Additionally,  we remark that a further strategy to increment the device gaps and Kondo temperature would be to use Pb superconducting islands \cite{kanne2021,Shen2025,microsoft2026}, which are characterized by much higher induced superconducting gaps, thus allowing for stronger tunneling couplings $t$ without violating the weak-tunneling assumptions.

Besides the CAR amplitude and the lead tunneling rate, also the role of the charging energy is crucial to obtain the charge Kondo effect in the poor man's tetron, as in all the other examples of charge Kondo effects. In our case, the electrostatic interactions in Eq. \eqref{charging} is necessary to obtain the effective two-level system playing the role of the quantum impurity in the Kondo effect. Therefore, $E_C$ must be strong enough that the ground states at $N_t=2,4$ are separated from excited states with even $N_t$ by a suitable gap. By comparing the energies of the ground states of the sectors with $N_t=0,6$ with the ones for $N_t=3$, we obtain similar gaps when $E_C \gtrsim \Delta_{\rm CAR}/6$, which sets the lower limit for the required value of the charging energy. Such requirement is easily fulfilled by double nanowire devices \cite{Vekris2022,Kurtossy2026}.
Below this threshold, instead, the system evolves towards a non-interacting regime, which can be approximately described as two uncorrelated two-dot systems analogous to the ones analyzed in the context of poor man's Majorana modes \cite{Liu2022}.

The perturbative expansion adopted to obtain the Hamiltonian \eqref{Kondo0} relies on the degeneracy of the ground states $\ket{2}$ and $\ket{4}$. When considering physical parameters away from the PH symmetric point, however, such degeneracy is generally split. We label this splitting with $h_0$ as it contributes to the Zeeman term $h$ in Eq. \eqref{Kondo2}. We have, in particular,
\begin{equation} \label{zeemanh}
h \approx h_0 + \sum_{\tau} \frac{\left(J_{\tau L}^2 -  J_{\tau R}^2\right)g_\Xi}{2\delta E^{(3)}}
\end{equation}
where $h_0$ is determined by the poor man's tetron $H_{\rm PMT}$, whereas the second term derives from the anisotropy of the lead couplings $J_\alpha$ as indicated in the second-order perturbative Hamiltonian in Eq. \eqref{Kondo2}.
The Zeeman energy $h$  drives the system away from the non-Fermi liquid point in a way captured by the parameter $T_h$ in Eq. \eqref{Th}, thus posing the main limitations for the observation of the charge Kondo effect. In order for the two-channel Kondo point to be observable, it is indeed necessary that $T_h \ll T,eV_b \ll T_K$. For experimental setups, the ideal condition is thus provided by the Zeeman temperature $T_h$ being comparable to the lowest achievable temperature $T^*$; therefore we target $T_h \lesssim T^* =20 $mK based on standard cryogenic setups. Such constraint imposes a maximum bound on the ground state splitting corresponding to 
\begin{equation} \label{constrh}
h < \sqrt{T_K T^*} \sim 5 \text{ \textmu eV } \sim \delta E^{(3)}/3\,. 
\end{equation}

In Fig. \ref{fig:GSSplitting} we illustrate the behavior of the bare splitting $h_0$ induced away from the PH symmetric point. In particular, we consider parameter regions where the previous constraint is fulfilled for symmetric lead couplings (thus $h_0 < \delta E^{(3)}/3$) and we  display the corresponding value of $h_0$. 
The central point of the central panel corresponds to the PH symmetric point. Away from it, the splitting $h_0$ vanishes along the dashed line, leading to an exact degeneracy  between the $|2\rangle$ and $|4\rangle$ ground states. We observe that there is a broad range of values of $\mu_{SC}$ around this degeneracy line where the constraint \eqref{constrh} is fulfilled and the gap $\delta E^{(3)}$ remains sizeable ($\delta E^{(3)}>0.01 \Delta \sim 20 \text{mK} $ for the parameters adopted in aluminum devices).

Within this parameter region with almost degenerate ground states $\ket{2}$ and $\ket{4}$, achieving the two-channel charge Kondo critical point described by Eq. \eqref{Kondo3} relies on two symmetries emerging at low energies: (i) the left-right mirror symmetry $\eta \leftrightarrow -\eta$ enforcing a vanishing contribution of the lead couplings to the Zeeman term (second term in the right hand side of Eq. \eqref{zeemanh}); (ii) the nanowire exchange symmetry $\tau \leftrightarrow -\tau$ that corresponds to the channel symmetry necessary for the onset of the non-Fermi behavior.

These symmetries are related to both the configurations of the couplings $J_a$ in Eq. \eqref{Hcoupling} and the possible nanowire dependence of the cotunneling and crossed Andreev reflection amplitudes in Eq. \eqref{hameff}. From the strong coupling analysis in the previous section, we know that the $\eta \leftrightarrow -\eta$ mirror symmetry plays a more important role, due to the RG relevance of the Zeeman term $h$. The channel symmetry, instead, drives the system away from the two-channel Kondo point only at second order. Therefore we focus on the constraints imposed by the violation of the left-right mirror symmetry, which, analogously to the ground state splitting, is captured by the temperature $T_h$ in Eq. \eqref{Th}. 

Also in this case, we can impose a constraint on the maximum anisotropy $\delta J_{LR} = \left|\sum_{\tau}\left(J_{\tau L} - J_{\tau R}\right)\right|/2$ to get $T_h < T^*$. When considering the PH symmetric point ($h_0=0$), we obtain:
\begin{equation}
\frac{\delta J_{LR}}{J} < \sqrt{\frac{T^*}{2\delta E^{(3)}}} \sim 20 \%\,. \label{tolerance}
\end{equation}
Due to the quadratic power in Eq. \eqref{Th}, therefore, we expect a considerable tolerance against the left/right anisotropy despite the Zeeman term $h$ being relevant. The result in Eq.  \eqref{tolerance} corresponds to a rough estimate obtained from Eq. \eqref{Th} by considering the strong coupling regime with $T_K \approx \delta E^{(3)} \approx J$, where $J$ represents the average lead-dot coupling. Concerning the less relevant channel anisotropy, the resulting constraints on the amplitudes $J_\alpha$ are less stringent than Eq. \eqref{tolerance}.  

As mentioned above, the ratio $R$ that establishes the system behavior at weak coupling is close to $1$ in the parameter regimes considered in the previous sections.  This stems from the assumption of PH symmetry and can be verified by truncating the perturbative expansions in Eqs. (\ref{Theta1} - \ref{Xi1}) to the lowest excited states (see Appendix \ref{app:Kondodet}). In Appendix \ref{app:gamma} we show that the contribution of the high energy states in the sectors with $N_t=3$ is indeed negligible and the ratio $R$ is close to $1$ for a broad range of physical parameters, relevant for the experimental realization of the poor man's tetron.

Finally, we observe that the key features of the poor man's tetron Hamiltonian $H_{\rm PMT}$ are stable under the introduction of more complex electrostatic interactions. In particular our model can be easily extended to include interdot and dot-island capacitances. In such scenario, it is still possible to identify a PH symmetric point characterized by the additional interactions $W_{\alpha\beta}(n_\alpha-n_{g,\alpha})(n_\beta-n_{g,\beta}) + U_{\alpha}(n_\alpha-n_{g,\alpha})(N-n_g)$. In this expression the electrostatic energies $W_{\alpha \beta}$ and $U_\alpha$ are derived by the capacitance matrix of the system. Particle-hole symmetry requires that the induced charges in the dots are set to $n_{g,\alpha}=1/2$. Under this tuning condition, the degeneracy of the sectors $(2,+)$ and $(4,+)$ is granted independently on the values and anisotropies of the $W$ and $U$ interactions. The related anisotropies may, however, still play a role in breaking the spatial mirror symmetries, thus causing perturbations to the states $\ket{2}$ and $\ket{4}$, and, consequently, to the Kondo Hamiltonian.

\section{Conclusions and outlook}

The poor man's tetron is an interacting quantum device designed to harness the high tunability of quantum dots for exploring exotic transport properties and observing emergent non-Fermi liquid critical phenomena. Besides the possibility of using it for the study of the topological Kondo effect \cite{Nitsch2025}, we showed in this work that it offers a controllable platform to observe a two-channel charge Kondo effect mediated by Cooper pairs.

This kind of charge Kondo effect has already been theoretically investigated for superconducting grains \cite{Garate2011} and hybrid superconductor-nanowire devices \cite{Pustilnik2017}. The poor man's tetron with fully polarized dots, however, allows for the onset of such effect exclusively based on spatial degrees of freedom, such that the incoming and outgoing Cooper pairs are formed by electrons in the four different quantum dots of this device.

In this respect, we showed that the crossed Andreev reflections within the two nanowires of the poor man's tetron dominate over other transport mechanisms in proximity of its particle-hole symmetric points and drive the system towards the related two-channel Kondo critical point at low temperatures. 

In this regime, the system behaves as a double Cooper pair splitter, with pairs of electrons entering the poor man's tetron from the two sides of one nanowire, and exiting from the two sides of the other [Fig. \ref{fig:GCAR}]. This is reflected in a quantized crossed Andreev conductance $G_{\rm CAR}= \frac{2e^2}{h}$ in the zero-temperature limit, with corrections linear in $T$ -- evidence of non-Fermi liquid behavior.

We estimate that in realistic devices with double nanowires proximitized with Al, the Kondo temperature  at strong coupling reaches 180mK, thus offering a broad range of the physical parameters for the study of this two-channel charge Kondo effect. As in the case of other two-channel Kondo models, anisotropies are detrimental for the observation of the non-Fermi liquid physics. From the scaling analysis of the main perturbations, however, we expect that the observation of the non-Fermi liquid properties of the system is resilient up to coupling anisotropies of order $20\%$, which can be easily achieved with current technology in superconducting double-nanowire systems \cite{Vekris2022,Kurtossy2026} (see also the gate-defined tetrons in Ref. \cite{microsoft2026}). Furthermore, the possibility of controlling via gate voltages the tunneling rates among both the external leads and the poor man's tetron and the inner components of the device offer the possibility of compensating disorder and other anisotropies of the system.

We thus envisage that the poor man's tetron offers a platform complementary to double quantum dots \cite{Potok2007} and fractional quantum Hall devices \cite{Iftikhar2015,Iftikhar2018} to observe the onset of non-Fermi liquid behavior and may potentially be exploited both as a source of correlated electrons and a building block for more complex quantum many-body systems.

\begin{acknowledgements}
The authors thank Jens Paaske and Bernard van Heck for useful discussions during the preliminary stage of this project and Virgil V. Baran for sharing his codes for the exact diagonalization of the poor man's tetron and reviewing a preliminary draft. Panel (a) of Fig. \ref{fig1} has been generated through the use of AI tools based on figures realized by Kasper Grove-Rasmussen and Rubén Seoane Souto for previous collaborations with some of the authors on hybrid double-nanowire devices. L.M. has been supported by the project "BALANCY" under the MSCA Seal of Excellence @Unipd programme.
\end{acknowledgements}

\appendix

\section{The effective Hamiltonian of the poor man's tetron} \label{app:pert}

This appendix summarizes the derivation of the effective Hamiltonian in Eq. \eqref{hameff}, based on the results in Refs. \cite{SoutoBaran,Nitsch2025}. For the sake of simplicity, we focus on the PH symmetric points, such that we consider the following values for the system parameters: $\mu_D=\mu_{SC}=0$ and $n_g=1$. More general results can be found in Ref. \cite{Nitsch2025}. Under the PH symmetry conditions, the unperturbed Hamiltonian reads:
\begin{equation}
H_0=E_C \left(N-1\right)^2 + \Delta\sum_{\tau}\left(c^{\dagger}_{\tau \uparrow }c^{\dagger}_{\tau \downarrow }\ee^{i\phi}+ {\rm H.c.}\right)\,.
\end{equation}

In the following we consider a perturbative regime $t \ll \Delta - E_C$ with $\Delta \gg E_C$, such that the tunneling amplitude is considered small compared with the other energy scales that characterize the system. In order to derive cotunneling and crossed Andreev amplitudes, we apply a second-order perturbative expansion to $H_{\rm PMT}$. The unperturbed Hamiltonian $H_0$ accounts for the proximity induced pairing in the central region of each nanowire, and, at the PH symmetric point ($\mu_{SC}=0$) can be easily diagonalized in terms of Bogoliubov modes that describe a pair of Andreev subgap states, with single-particle Hamiltonian:
\begin{equation}
H_{SC} = \sum_{s,\tau} \Delta f^\dag_{\tau s} f_{\tau s}\,.
\end{equation}
Here the Bogoliubov modes are defined by:
\begin{align} \label{Bog1}
&f_{\tau \Up } = \frac{1}{\sqrt{2}}\left( c_{\tau \Up }\ee^{-i\varphi/2} +  c^\dag_{\tau \Dn }\ee^{i\varphi/2}\right),\\
&f_{\tau \Dn } = \frac{1}{\sqrt{2}}\left( c_{\tau \Dn }\ee^{-i\varphi/2} -  c^\dag_{\tau \Up }\ee^{i\varphi/2}\right),\label{Bog2}
\end{align}
such that:
\begin{equation} 
c^\dag_{\tau s } = \frac{1}{\sqrt{2}}\left(  f^\dag_{\tau s } -s  f_{\tau \bar{s} }\right)\ee^{-i\varphi/2} \,,
\end{equation}
where $s=\pm 1$ indicates spin up and down respectively and $\bar{s}$ labels a spin aligned opposite to $s$. As a function of the Bogoliubov modes $f_s$, the tunneling terms in Eq. \eqref{hamsp} read:
\begin{widetext}
\begin{equation}
H_t = 
- \frac{t}{\sqrt{2}} \sum_{\tau,\eta} \left[\cos\alpha \left( f^\dag_{\tau\Dn} +   f_{\tau\Up}\right) -\eta \sin \alpha \left( f^\dag_{\tau\Up} -   f_{\tau\Dn}\right) \right] d_{\tau \eta}\ee^{-i \varphi/2} + {\rm H.c.}.
\end{equation}
The operator $\ee^{-i\varphi/2}$ suitably increases the charge of the SC island: it is conjugate to $N$ and obeys the commutation relation $\left[N,\ee^{-i\varphi/2}\right]=\ee^{-i\varphi/2}$. 

Next, we consider a regime of strong induced pairing, such that $E_C<\Delta$ and the lowest energy states are characterized by unoccupied Bogoliubov modes. Additionally, we observe that the spin-orbit coupling can be described in terms of a spin rotation $\ee^{-i\eta \alpha \sigma_y}$ by $-\eta \alpha$ along the $y$ axis, starting from the spin down polarization of the quantum dots (assuming a strong out of plane magnetic field); therefore, the elastic cotunneling processes at second order can be conveniently written as \cite{Nitsch2025}:
\begin{multline} \label{cot1}
H_{\rm COT} = \sum_{\tau, \eta^{\prime}, \eta^{\prime\prime}, s} -d^\dag_{\tau \eta^{\prime \prime}} \left(\ee^{i \eta^{\prime \prime} \alpha \sigma_y}\right)_{\Dn s} \frac{t^2}{2\left[\Delta + E_C\left(2N-1\right)\right]}\left(\ee^{-i \eta^{\prime } \alpha \sigma_y}\right)_{s\Dn}d_{ \tau \eta^{\prime } } \\
-  d_{  \tau \eta^{\prime \prime}} \left(\ee^{-i \eta^{\prime \prime} \alpha \sigma_y}\right)_{s\Dn} \frac{t^2}{2\left[\Delta + E_C\left(3-2N\right)\right]}\left(\ee^{i \eta^{\prime } \alpha \sigma_y}\right)_{\Dn s}d^\dag_{\tau \eta^{\prime }}\,,
\end{multline}
\end{widetext}
where $N$ is the total charge of the superconducting island in the initial state. Eq. \eqref{cot1} includes both terms with $\eta' \neq \eta^{\prime \prime}$, which correspond to the elastic cotunneling events, and terms with $\eta' = \eta^{\prime \prime}$ which describe tunnelings back and forth of an electron between a quantum dot and the superconducting region. The latter yields charge-dependent corrections to the dot energy levels.

From the previous expression it is easy to derive the cotunneling terms. The related amplitudes for the values of the charge $N=0$ and $N=2$ are expressed in Eq. \eqref{tcot} and determine the dynamics of the most relevant sectors for the definition of the charge Kondo Hamiltonian. More general expressions for the cotunneling amplitudes away from the charge degeneracy point can be found in Refs. \cite{SoutoBaran,Nitsch2025}.

The terms of Eq. \eqref{cot1} with $\eta^{\prime}= \eta^{\prime\prime}$, instead, give rise to shifts of the energy levels of the dots which are suitably expressed as:
\begin{equation} \label{hamshift}
H_{\rm SHIFT} = \sum_{\tau,\eta} \tilde{\mu}_+(N) d^\dag_{\tau \eta} d_{\tau \eta}  + \tilde{\mu}_-(N) d_{\tau \eta} d^\dag_{\tau \eta}\,,
\end{equation}
where:
\begin{align}\label{mutildep}
&\tilde{\mu}_+(N) = -\frac{t^2}{2\left[\Delta + E_C\left(2N-1\right)\right]} \,, \\
&\tilde{\mu}_-(N)=-\frac{t^2}{2\left[\Delta + E_C\left(3-2N\right)\right]}\,.\label{mutildem}
\end{align}
These expressions make it evident that the terms in $H_{\rm SHIFT}$ are PH-symmetric and, in particular, we have:
\begin{equation}
\tilde{\mu}_+(N) = \tilde{\mu}_-(2-N)\,.
\end{equation}

The calculation of the crossed Andreev reflection amplitude is performed in an analogous way. When considering transitions between states with $N$ and $N+2$ electrons in the superconducting island, we obtain an effective p-wave pairing between the polarized electrons in the quantum dots of the form:
\begin{widetext}
\begin{equation}\label{carH}
H_{\rm CAR}(N) = \frac{t^2}{2} \ee^{-i\varphi} \sum_{\tau, \eta^{\prime}, \eta^{\prime\prime}, s} d_{\tau \eta^{\prime \prime}}\left(\ee^{-i \eta^{\prime \prime} \alpha \sigma_y}\right)_{\bar{s}\Dn} \frac{s}{\Delta - E_C}\left(\ee^{-i \eta^{\prime } \alpha \sigma_y}\right)_{s\Dn} d_{\tau \eta^\prime}  + {\rm H.c.}\,,
\end{equation}
\end{widetext}
where $s=\pm 1$ labels the spin of the virtual Bogoliubov state involved in the process and we consider transition between the degenerate states with charge $N=0$ and $N=2$. The same expression, however holds also for the transitions between non-degenerate states with different charges in regimes where $\Delta \gg t \gtrsim E_C$. The previous equation results into the crossed Andreev amplitude expressed in Eq. \eqref{deltacar}.

\section{Symmetry sectors and eigenstates} \label{app:sectors}

In order to derive the charge Kondo Hamiltonian, it is useful to consider the structure of the symmetry blocks $\left(N_t,P_\Dnt\right)$ at the charge degeneracy point. The perturbative calculation in Sec. \ref{sec:Kondo} relies indeed on the form of the eigenstates of $H_{\rm eff}$. In the following, we label the basis states as $\ket{n_{\Upt L} n_{\Upt R} n_{\Dnt L} n_{\Dnt R}, N}$ a function of the dot occupation numbers and the charge $N$.

\begin{widetext}
\subsection{Even sector $(N_t=2m,P_\Dnt=+)$} \label{App:evenp}
This sector corresponds to the parities $P_\Upt=1,\,P_\Dnt=1$.

Its basis is $\left\{\ket{0000,N_t},\ket{0011,N_t-2},\ket{1100,N_t-2},\ket{1111,N_t-4}\right\}$ and the related Hamiltonian blocks acquire the form:
\begin{equation} \label{Heven1}
H_{\rm eff}^{\text{even}+} =  \begin{pmatrix} 
[4 \tilde\mu_- + H_c](N_t)& -\Delta_{\rm CAR} & -\Delta_{\rm CAR} & 0 \\
-\Delta_{\rm CAR} & [2\tilde{\mu}_+ + 2\tilde{\mu}_- +H_c](N_t-2) & 0 & -\Delta_{\rm CAR} \\
-\Delta_{\rm CAR} & 0 & [2\tilde{\mu}_+ + 2\tilde{\mu}_- +H_c](N_t-2) & -\Delta_{\rm CAR} \\
0 & -\Delta_{\rm CAR} &-\Delta_{\rm CAR} & [4\tilde{\mu}_+ +H_c](N_t-4)
\end{pmatrix}
\end{equation}

For $N_t=4$, at the charge degeneracy point $n_g=1,\, \mu=0$, the charging energy contribution to the diagonal entries of $H_{\rm eff}^{\text{even}+}$ become $9E_C,E_C,E_C,E_C$. For sufficiently strong interactions $E_C > \Delta_{\rm CAR},\, -\tilde{\mu}_{\pm}(2)$, the state $\ket{0000,4}$ acquires a considerably higher energy than the others and can be effectively removed from the low-energy basis. The Hamiltonian can then be approximated as:
\begin{equation}\label{Hevenred}
H_{\rm eff}^{\text{even}+} (N_t=4)\xrightarrow{E_C \gg \Delta_{\rm CAR}} \begin{pmatrix} 
2\tilde{\mu}_+(2)+2\tilde{\mu}_-(2) +E_C & 0 & -\Delta_{\rm CAR} \\
 0 & 2\tilde{\mu}_+(2)+2\tilde{\mu}_-(2) + E_C & -\Delta_{\rm CAR} \\
 -\Delta_{\rm CAR} &-\Delta_{\rm CAR} & 4\tilde{\mu}_-(2) + E_C
\end{pmatrix}.
\end{equation}  

In this limit, the ground state energy results:
\begin{multline} 
E_{\rm GS}^{(N_t=2,4; +)} \approx E_C+3\tilde{\mu}_-(2) + \tilde{\mu}_+(2) - \sqrt{2 \Delta_{\rm CAR}^2 + \left(\tilde{\mu}_+(2)-\tilde{\mu}_-(2)\right)^2} \\
= E_C - \frac{t^2}{\Delta - E_C}\left[ \frac{2\Delta + 4E_C}{\Delta + 3 E_C} + \sqrt{2 \sin^2 (2\alpha) + \frac{4E_C^2}{\left(\Delta + 3E_C \right)^2}} \right],
\end{multline}

where this expression holds for both $N_t=2$ and $N_t=4$ due to the PH symmetry. The $N_t=2$ sector is indeed analogous: in this case the state $\ket{1111,-2}$ acquires a higher energy and becomes decoupled for large $E_C$. 

Based on the approximation \eqref{Hevenred}, the wavefunction of the ground state for $N_t=4$ can be approximated as:
\begin{equation} \label{ket4}
\ket{4} \approx \frac{\sin{\upchi}}{\sqrt{2}}\left(\ket{0011,2}+\ket{1100,2} \right)+\cos{\upchi}\ket{1111,0}
\end{equation}
with the parameter $\upchi$ corresponding to:
\begin{equation}
\upchi=\frac{1}{2}{\rm Arg}\left[\tilde\mu_{+}(2)-\tilde\mu_-(2) +i\sqrt{2} \Delta_{\rm CAR}\right]\,.
\end{equation}
In a regime with $\Delta \gg E_C$ and $\alpha \approx \pi/4$, $\upchi \lesssim \pi/4$.
With analogous approximation, we find that the ground state of the $(2,+)$ sector can be expressed as:
\begin{equation} \label{ket2}
\ket{2} \approx \frac{\sin{\upchi}}{\sqrt{2}}\left(\ket{0011,0}+\ket{1100,0} \right)+\cos{\upchi}\ket{0000,2}\,.
\end{equation}

\subsection{Even sector $(N_t=2m,P_\Dnt=-)$}

This sector corresponds to $\left(P_\Upt=-1,P_\Dnt=-1\right)$ and its basis displays only states having two electrons in the external dots and charge $N= N_t-2$ in the superconducting island: $\left\{\ket{0101,N_t-2},\ket{0110,N_t-2},\ket{1001,N_t-2},\ket{1010,N_t-2}\right\}$. The effective Hamiltonian reads:
\begin{equation}
H_{\rm eff}^{\text{even}-} = [2\tilde{\mu}_+ + 2\tilde{\mu}_- +H_c](N_t-2)\Id +\begin{pmatrix} 
0 & -t_{\rm COT}(N_t-2) & -t_{\rm COT}(N_t-2) & 0 \\
-t_{\rm COT}(N_t-2) & 0 & 0 & -t_{\rm COT}(N_t-2) \\
-t_{\rm COT}(N_t-2) & 0 & 0 & -t_{\rm COT}(N_t-2) \\
0 & -t_{\rm COT}(N_t-2) &-t_{\rm COT}(N_t-2) & 0
\end{pmatrix}
\end{equation}
where $\Id$ labels a $4\times 4$ identity matrix. For $N_t=2$ and $N_t=4$, the ground state energy reads:
\begin{equation}
E_{\rm GS}^{(N_t=2,4; -)} = E_C+2\tilde{\mu}_+(2) + 2\tilde{\mu}_-(2)   - 2|t_{\rm COT}(0)| = 
E_C - \frac{2t^2\left[\Delta + E_C\left(1 + 2 \cos(2\alpha)  \right)\right]}{\left(\Delta + 3E_C \right)\left(\Delta -E_C \right)}
\end{equation}

\subsection{Odd sectors $(N_t=2m+1,P_\Dnt=\pm)$ }

Let us consider first the sector with $P_\Upt=-1,\, P_\Dnt=1$; the other is symmetric under the exchange $\tau \to -\tau$.

The basis of the $(2m+1,+)$ sector is given by $\left\{\ket{0100,N_t-1},\ket{1000,N_t-1},\ket{0111,N_t-3},\ket{1011,N_t-3}\right\}$. Its effective Hamiltonian reads:
\begin{equation}
H_{\rm eff}^{\text{odd}+} = \begin{pmatrix} 
[\tilde{\mu}_+ + 3\tilde{\mu}_- +H_c](N_t-1) & -t_{\rm COT}(N_t-1) & -\Delta_{\rm CAR} & 0 \\
-t_{\rm COT}(N_t-1) & [\tilde{\mu}_+ + 3\tilde{\mu}_- +H_c](N_t-1) & 0 & -\Delta_{\rm CAR} \\
-\Delta_{\rm CAR} & 0 & [3\tilde{\mu}_+ + \tilde{\mu}_- +H_c](N_t-3) & -t_{\rm COT}(N_t-3) \\
0 & -\Delta_{\rm CAR} &-t_{\rm COT}(N_t-3) & [3\tilde{\mu}_+ + \tilde{\mu}_- +H_c](N_t-3)
\end{pmatrix}
\end{equation}

We consider in particular the sector with $N_t=3$, which provides the virtual states that mediate the interactions between the ground states with $N_t=2,4$ in the Kondo Hamiltonian \eqref{Kondo0}. At the charge degeneracy point all the diagonal elements acquire the same value $E_C + \tilde{\mu}_+(2) + 3 \tilde{\mu}_-(2)$.
The ground states of these sectors are thus doubly degenerate and their energy is:
\begin{multline} 
E_{GS}^{(N_t=3;\pm)} = E_C + \tilde{\mu}_+(2) + 3 \tilde{\mu}_-(2) - \sqrt{t_{\rm COT}^2(2) + \Delta_{\rm CAR}^2}\\
= E_C - \frac{t^2}{\Delta - E_C}\left[ \frac{2\Delta + 4E_C}{\Delta + 3 E_C} + \sqrt{ \sin^2 (2\alpha) + \frac{4\cos^2(2\alpha)E_C^2}{\left(\Delta + 3E_C \right)^2}} \right]\,,
\end{multline}
such that $E_{GS}^{N_t=2,4;+}< E_{GS}^{(N_t=3;\pm)}< E_{GS}^{N_t=2,4;-}$.

In particular the gap between the $N_t=3$ ground states and the global ground states can be approximated as:
\begin{equation} \label{dE3}
\delta E^{(3)}\equiv E_{GS}^{(N_t=3;\pm)} - E_{GS}^{(N_t=2,4;+)}  \approx \sqrt{2\Delta_{\rm CAR}^2 + \left(\tilde\mu_+(2) - \tilde\mu_-(2) \right)^2} - \sqrt{\Delta_{\rm CAR}^2 + \cos^2(2\alpha) \left(\tilde\mu_+(2) - \tilde\mu_-(2) \right)^2}\,,
\end{equation}
where this relation is derived from the approximated Hamiltonian \eqref{Hevenred} and holds for $E_C > \Delta_{\rm CAR}, -\tilde{\mu}_{\pm}(2)$.

The ground states of the $(3,+)$ sectors can be written as:
\begin{align} \label{o1b}
&\ket{3,+,1}= \frac{1}{\sqrt{2}}\left[\cos\uptheta \ket{0100,2} - \cos\uptheta \ket{1000,2} + \sin\uptheta  \ket{0111,0} -\sin\uptheta \ket{1011,0} \right]\,,\\
&\ket{3,+,2}= \frac{1}{\sqrt{2}}\left[\sin\uptheta \ket{0100,2} + \sin\uptheta \ket{1000,2} + \cos\uptheta  \ket{0111,0} +\cos\uptheta \ket{1011,0} \right] \label{o1c}\,,
\end{align} 
\end{widetext}

where the parameter $\uptheta$ is defined as:
\begin{equation} \label{arg}
\uptheta=\frac{1}{2}{\rm Arg}\left[t_{\rm COT}(0) +i \Delta_{\rm CAR}\right]\,.
\end{equation}
In the perturbative regime $\Delta \gg t,E_C$, the angle $\uptheta$ approaches $\pi/4$ as $\Delta_{\rm CAR} \gg t_{\rm COT}(0)$. 

The sector $(3,-)$ displays exactly the same behavior of $(3,+)$ based on the mapping $\tau \to -\tau$, and it is characterized by analogous two-fold degenerate eigenstates.

\section{Second-order calculation of the Kondo Hamiltonian} \label{app:Kondodet}

The second-order perturbation theory adopted to derive the Hamiltonian $H_K$ in Eq. \eqref{Kondo0} relies on the assumption of small tunnelling amplitudes between leads and poor man's tetron, namely $J_{\tau\eta} \ll \delta E^{(3)}$. Since the main contribution to $\delta E^{(3)}$ at the PH symmetric points is provided by $\Delta_{\rm CAR}$ (for sizeable $\alpha$), the perturbative Hamiltonian \eqref{Kondo0} is valid in a regime such that:  $J_{\tau\eta} \ll \Delta_{\rm CAR}, E_C, t \ll \Delta$.

The $\Theta$ terms in Eqs. \eqref{Kondo0} and \eqref{Theta1} are proportional to the $S^{-}$ operator as they correspond to a pair of electrons moving from the poor man's tetron to the leads. In particular, $\Theta^{\alpha \beta}$ acquires the following antisymmetric form:
\begin{multline} \label{Theta2}
\Theta^{\alpha \beta} = \frac{g_\Theta}{\delta E^{(3)}}\left( \delta_{\alpha,1}\delta_{\beta,2}-\delta_{\beta,1}\delta_{\alpha,2} \right.
\\
\left. 
+\delta_{\alpha,3}\delta_{\beta,4}-\delta_{\beta,3}\delta_{\alpha,4}\right)|2\rangle\langle4|\,.
\end{multline}
Therefore, $\Theta$ describes pairing processes among electrons belonging to the two leads connected with the same nanowire. In this expression, we adopted for convenience $\delta E^{(3)}$ as energy scale, and we accordingly introduced the numerical factor $g_\Theta$, which depends on the wavefunctions of the ground states $\ket{2}$ and $\ket{4}$ as well as the eigenstates of the $N_t=3$ sectors.

To gain an approximate estimate of $g_\Theta$, we can limit the sum over the $N_t=3$ states in Eq. \eqref{Theta1} to the lowest lying states of the $(3,\pm)$, thus truncating $n$ to the first two values, $n=1,2$, which correspond to the eigenstates in Eqs. (\ref{o1b},\ref{o1c}). This approximation is justified by the fact that, based on Eq. \eqref{dE3}, $\delta E^{(3)} \approx (\sqrt{2}-1)\Delta_{\rm CAR}$ whereas the gap between the eigenstates of $H_{\rm eff}^{\rm odd+}$ is about $2\Delta_{\rm CAR}$, thus approximately 5 times larger. Additionally, to obtain an approximate analytical form for $g_\Theta$, we also consider the strong coupling regime $\Delta_{\rm CAR} < E_C$, such that we apply the approximations in Eqs. (\ref{ket4},\ref{ket2}). Under these assumptions, we obtain:
\begin{widetext}
\begin{equation}
g_\Theta \approx \frac{2\sqrt{2}\sin(2\upchi) + \left(3+\cos(2\upchi)\right)\sin(2\uptheta)}{8} \,\xrightarrow{\Delta_{\rm CAR} \gg t_{\rm COT}(0)}\, \frac{2\sqrt{2} + 3}{8}\,. \label{gthetaa}
\end{equation}
The final rough approximation for $\Delta_{\rm CAR} \gg t_{\rm COT}(0)$ (thus $\upchi,\uptheta \to \pi/4$) provides a very good approximation for a broad range of $\alpha$ and $E_C$.

A similar analysis can be performed for the operators $\Upsilon$ and $\Xi$. In particular, their general form is a symmetric matrix of the kind:

\begin{align} \label{Upsilon2}
& \Upsilon^{\alpha \beta} = - \left[\frac{g_{\Xi}}{\delta E^{(3)}} \delta_{\alpha \beta} + \frac{g_{\rm cot}}{\delta E^{(3)}}\left(\delta_{\alpha,1}\delta_{\beta,2}+\delta_{\alpha,2}\delta_{\beta,1}+\delta_{\alpha,3}\delta_{\beta,4}+\delta_{\alpha,4}\delta_{\beta,3}\right) \right] \ket{4} \bra{4}\,,\\
&\Xi^{\alpha \beta} = - \left[\frac{g_{\Xi}}{\delta E^{(3)}} \delta_{\alpha \beta} + \frac{g_{\rm cot}}{\delta E^{(3)}}\left(\delta_{\alpha,1}\delta_{\beta,2}+\delta_{\alpha,2}\delta_{\beta,1}+\delta_{\alpha,3}\delta_{\beta,4}+\delta_{\alpha,4}\delta_{\beta,3}\right) \right] \ket{2} \bra{2}\,. \label{Xi2}
\end{align}
In this case, we introduced two numerical factors: $g_\Xi$ is responsible for the $V$ terms in Eq. \eqref{Kondo1}; $g_{\rm cot}$, instead, determines the amplitudes $J^{\rm (cot)}_\tau$.
Also in this case, we can obtain approximate expressions of these constants by truncating the $n$ summation in Eqs. (\ref{Upsilon1},\ref{Xi1}) and considering the strong coupling regime $E_C > \Delta_{\rm CAR}$. We get:
\begin{align}
&g_\Xi \approx \frac{3+ \cos(2\upchi)+ 2\sqrt{2}\sin(2\upchi)\sin(2\uptheta)}{8} \,\xrightarrow{\Delta_{\rm CAR} \gg t_{\rm COT}(0)} \,\frac{2\sqrt{2} + 3}{8}\,, \label{gxia}\\
&g_{\rm cot} \approx -\frac{\left[1+ 3\cos(2\upchi) \right]\cos(2\uptheta)}{8} \, \xrightarrow{\Delta_{\rm CAR} \gg t_{\rm COT}(0)} \,0\,.
\end{align}
\end{widetext}

Therefore, in our regime of interest, the coefficients $g_\Xi$ and $g_\Theta$ have comparable amplitudes $(\sim 0.73)$,  whereas the cotunneling parameter $g_{\rm cot}$, thus the amplitudes $J^{\rm(cot)}$ are much weaker, reflecting the energy hierarchy $\Delta_{\rm CAR} \gg t_{\rm COT}(0)$.

When considering the isotropic case with equal $J_{\tau\eta}=J$, this results in the following amplitudes for the Hamiltonian \eqref{Kondo2}:
\begin{align}
&J^{(\perp)}_\tau = \frac{2g_\Theta J^2}{\delta E^{(3)}}\,, \label{Perp_0} \\
&J^{(z)}_\tau = \frac{g_\Xi J^2}{\delta E^{(3)}}\,, \label{Zeta_0} \\
&J_\tau^{\rm(cot)} = \frac{g_{\rm cot} J^2}{\delta E^{(3)}}\,,\\
&W_\tau =h =0\,.
\end{align}
The last terms vanish as they are proportional to $\sum_{\tau} \left(J_{\tau L}^2 -  J_{\tau R}^2\right)g_\Xi/\delta E^{(3)}$. From these estimates we obtain that the ratio $R\equiv {J^{(\perp)}}/{(2J^{(z)})} = g_\Theta / g_\Xi \approx 1$, which indicates that the bare values of the couplings lie close to the red line in Fig. \ref{fig:RGFlow}. 
We also observe that the previous equations allow us to express the perturbative parameter $\rho J^{(\perp)}$ as
\begin{equation}
\rho J^{(\perp)} = 2g_\Theta \frac{\rho J^2}{\delta E^{(3)}} = \frac{g_\Theta}{\pi}\frac{\Gamma_e}{\delta E^{(3)}} \approx 0.23 \frac{\Gamma_e}{\delta E^{(3)}}
\end{equation}
as a function of the lead-dot tunneling rate $\Gamma_e \equiv 2\pi \rho J^2$.

\section{Detail on the poor man's renormalization of the Kondo couplings} \label{app:RG}
The poor man's scaling procedure \cite{Anderson1, Anderson2} allows us to determine the scaling of the coupling constants in the weak coupling regime, as presented in the main text. The bandwidth of the system is iteratively reduced $D\rightarrow D-\delta D$ with $\delta D>0$ by integrating out virtual high energy states at second order in perturbation theory, which results in a flow of the couplings. The RG equations (\ref{RGz},\ref{RGperp}) for the Kondo couplings $J_{\tau}^{(z)}$ and $J^{(\perp)}_{\tau}$ are straightforward to obtain, and we refer to Ref. \cite{Hewson_1993} for a detailed analysis. In particular, the exchange interaction associated with $J_{\tau}^{(\perp)}$ renormalizes the values of the couplings $V_{\tau L}$ and $V_{\tau R}$ in Eqs. (\ref{Kondo1},\ref{V1}) through second order processes, in which a pair of electrons enters the device from two opposite leads, occupy an intermediate excited state, and subsequently tunnels back into the same leads, thereby leaving the total charge of the tetron unchanged. The correction to $V_{\tau L}$ and $V_{\tau R}$ is the same, and is given by: 
\begin{equation}
   \delta V_{\tau L}=  \delta V_{\tau R}=\rho \left(J_{\tau}^{(\perp)}\right)^2\frac{\delta D}{D},
\end{equation}
where $\rho$ represents the density of states of the leads at the Fermi level. Therefore, the scaling equation for $J_{\tau}^{(z)}=\left(V_{\tau L} + V_{\tau R}\right)/2$ reads:
\begin{equation}
     \frac{ {\rm d}J_\tau^{(z)}}{{\rm d} l} = \rho \left(J^{(\perp)}_\tau\right)^2,
\end{equation}
where we introduced the RG parameter $l=\ln(D_{0}/D)$. Here, $D_{0}$ represents the initial cutoff scale, corresponding to the energy of the highest excitations of the bare poor man's tetron Hamiltonian. 

Conversely, the cotunneling terms associated with $J^{(\rm{cot})}_{\tau}$ do not renormalize at second order. In principle, an effective cotunneling interaction could be generated, for instance, via a crossed Andreev process, where a pair of electrons tunnel from the device into two opposite leads, followed by a local Andreev process, in which a pair of electrons from the same lead tunnel into the poor man's tetron. However, these local Andreev processes are suppressed in the low energy sector of the model, due to the assumption of fully spin-polarized dots. Therefore, the coupling $J^{(\rm{cot})}_{\tau}$ remains marginal at second order, namely:
\begin{equation}
    \frac{{\rm d} J^{(\rm{cot})}_\tau}{{\rm d}l}=0.
\end{equation}

These considerations yield the equations (\ref{RGz},\ref{RGperp},\ref{RGcot}), which describe the RG flow of the Cooper pair charge Kondo Hamiltonian (\ref{Kondo3}) in a weak coupling regime and for left/right symmetric couplings ($W_{\tau}=h=0$). The additional third order corrections to $J^{(z)}_\tau$ and $J^{(\perp)}_\tau$ can also be obtained through standard techniques (see Ref. \cite{Garate2011} for the analogous equations in the context of superconducting grains). 

The solutions of the second-order RG equations are conveniently derived considering two different cases, depending on the sign of the invariant $\mathcal{I}$. If $\mathcal{I}>0$, namely if $J^{(\perp)}>2J^{(z)}$, the solutions read:
\begin{equation}
J^{(\perp)}\left(l\right)=\frac{\mathcal{I}^{1/2}}{\sin\left[2\rho \mathcal{I}^{1/2}\left(\bar{l} - l\right)\right]}, \label{4.92}
\end{equation}
\begin{equation}J^{(z)}\left(l\right)=\frac{\mathcal{I}^{1/2}}{2} \cot\left[2\rho \mathcal{I}^{1/2} \left(\bar{l} - l\right)\right], \label{4.93}
\end{equation}
where $\bar{l}$ denotes the RG scale at which the couplings diverge, given by: 
\begin{equation}
\bar{l} = \frac{1}{2\rho \mathcal{I}^{1/2}}\arccos \frac{2J^{(z)}(0)}{J^{(\perp)}(0)}.
\end{equation}
\begin{figure*}[t!]
    \centering
    \includegraphics[width=\textwidth]{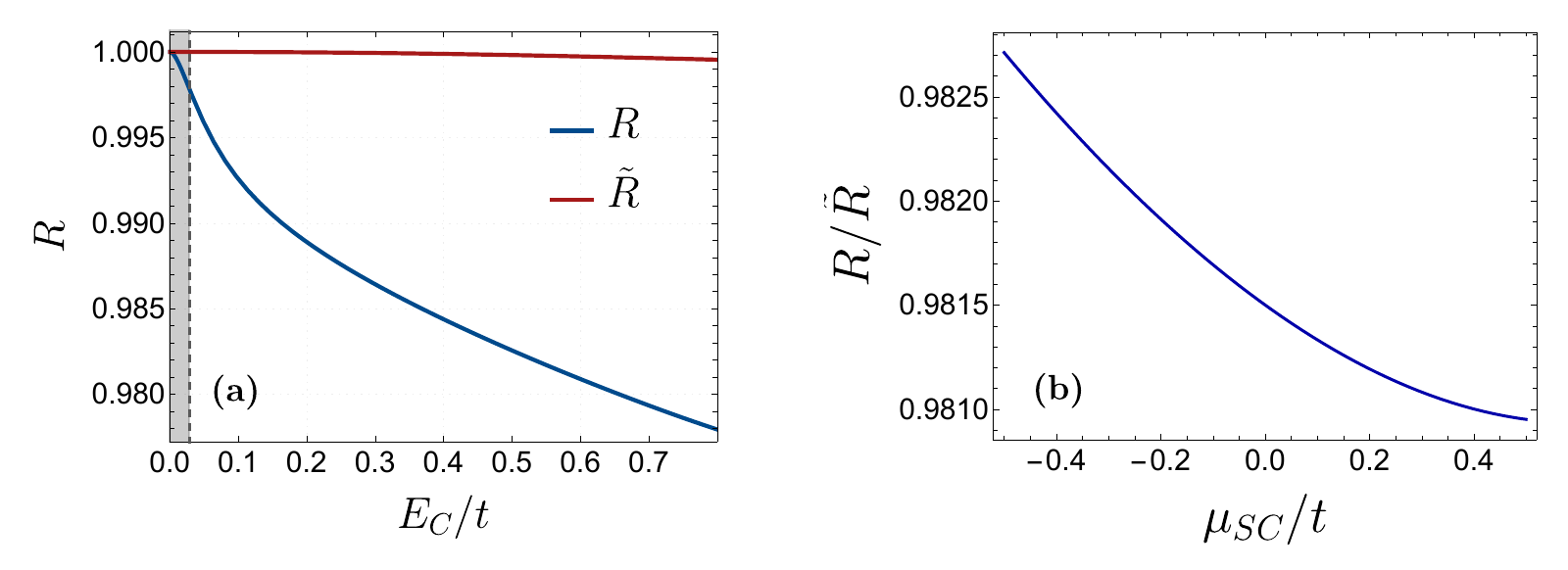}
    \caption{(a) Values of the ratios $\tilde R$ and $R$ as a function of $E_C$ for $\Delta=5t, \; \mu_{SC}=0, \; n_g=1, \; \mu=0, \; \alpha=\pi/6$. We observe that the dependence from the charging energy is very mild within the depicted range of electrostatic interactions. The shaded area at low $E_C$ indicates the regime $E_C<\Delta_{\text{CAR}}/6$ which violates the two-state approximation for realizing a Kondo model. (b) Ratio $R/\tilde{R}$ as a function of $\mu_{SC}$. The ratio is computed on the degeneracy line (dashed line in Fig. \ref{fig:GSSplitting}), where the ground states of the $(2,+)$ and $(4,+)$ sectors are degenerate, and for $\Delta=4t, \; E_C=0.5t, \; n_g=1, \;  \alpha=\pi/6$. We observe that the approximation $\tilde{R}$ provides a good estimate for the ratio $R$.}
    \label{fig:ECplot}
\end{figure*}
An estimate of the Kondo temperature $T_{K}$ can be obtained from $\ln(D_{0}/T_{K})=\bar{l}$. 
In our analysis, we derived the Kondo Hamiltonian by considering second-order processes involving the $N_{t}=3$ sectors. If we disregard the contributions from higher energy states, the initial cutoff scale $D_{0}$ can be approximated with the energy gap $\delta E^{(3)}$, yielding: 
\begin{equation}
T_K \simeq \delta E^{(3)}\ee^{-\frac{1}{2\rho \mathcal{I}^{1/2}}\arccos \frac{g_\Xi}{g_\Theta}},
\end{equation}
where we substituted the initial values of the couplings in (\ref{Perp_0}, \ref{Zeta_0}).

For $\mathcal{I}<0$, i.e. $J^{(\perp)}<2J^{(z)}$, we have instead:
\begin{equation}
J^{(\perp)}\left(l\right)=\frac{|\mathcal{I}|^{1/2}}{\sinh\left[2\rho |\mathcal{I}|^{1/2}\left(\bar{l} - l\right)\right]}, \label{Jperp2}
\end{equation}
\begin{equation}
    J^{(z)}\left(l\right)=\frac{|\mathcal{I}|^{1/2}}{2} \coth\left[2\rho |\mathcal{I}|^{1/2} \left(\bar{l} - l\right)\right],
\end{equation}
such that:
\begin{equation}
\bar{l} = \frac{1}{2\rho |\mathcal{I}|^{1/2}}\text{arccosh} \frac{2J^{(z)}(0)}{J^{(\perp)}(0)},
\end{equation}
and 
\begin{equation}
T_K \simeq \delta E^{(3)}\ee^{-\frac{1}{2\rho \mathcal{I}^{1/2}}\text{arccosh} \frac{g_\Xi}{g_\Theta}}.
\end{equation}

Within our second order analysis, both the Kondo coupling constants $J^{(\perp)}$ and $J^{(z)}$ diverge as an effect of the renormalization flow. However, by including also the third order correction, one recovers the standard 2-channel Kondo fixed point at finite couplings \cite{Iftikhar2015,Vojta2006}. 

In the presence of an asymmetry between the left and right couplings, both the Zeeman term $h$ and the $W$ perturbation are initially different from zero. Equation (\ref{RGh}) for the Zeeman field is obtained simply from scaling arguments, while $W$ is marginal at second order (\ref{RGW}). Indeed, $W$ is proportional to $V_{\tau L}-V_{\tau R}$ and, following the discussion presented before, we have:
\begin{equation}
    \delta W_{\tau}=0.
\end{equation}

Let us estimate now the Andreev conductance in Eq. \eqref{CARG} at zero voltage bias for $T_K < T < \delta E^{(3)}$, which is the range of validity of our RG analysis for the poor man's tetron.

The Kondo conductance can be estimated through the conductance \eqref{CARG} by replacing the coupling constant $J^{(\perp)}$ with its renormalized value obtained when the bandwidth $D$ of the lead electrons is reduced to the same level of the temperature $T \sim D < D_0$, such that $l = -\ln \left(T/D_0\right)$ \cite{Pustilnik2017}. 
Based on the solution \eqref{4.92} of the RG equations in the weak coupling regime, the interwire conductance for the renormalized coupling $J^{(\perp)}(l)$ reads:
\begin{equation} \label{CARG2}
G_{\rm CAR}= 2\tilde{G}_{0}\int d\epsilon \left(-\partial_{\epsilon} n_{F}\right)\frac{\rho^2\mathcal{I}}{\sin^2\left[2\rho \mathcal{I}^{1/2}\ln (T/T_K) \right]}\,,
\end{equation}
where we used the relation $T_K=\delta E^{(3)} \ee^{-\bar{l}}$.

In the weak coupling limit, where $\mathcal{I}^{1/2}\rho \ll 1$, and for $T_K < T < \delta E^{(3)}$, we can approximate Eq. \eqref{CARG2} as
\begin{equation}
G_{\rm CAR} \sim \frac{\tilde{G}_{0}}{2} \int d\epsilon \left(-\partial_{\epsilon} n_{F}\right) \left[\ln (T/T_K)\right]^{-2}, 
\end{equation}
which matches Eq. \eqref{GCART} and is consistent with the result obtained by estimating the conductance with a rate equation approach by considering the renormalized sequential tunneling contribution \cite{Matveev1995}. The same result is obtained also in the regime $J^{(\perp)}<2J^{(z)}$ by using Eq. \eqref{Jperp2}.

\section{Ratio $J^{(\perp)}/(2J^{(z)})$ as a function of the system parameters} \label{app:gamma}
In Sec. \ref{sec:experiments}, we introduced the parameter $R\equiv J^{(\perp)}/(2J^{(z)})= g_{\Theta}/g_{\Xi}$, where $g_{\Theta}$ and $g_\Xi$ are defined through the Kondo interactions in Eqs. (\ref{Theta2}, \ref{Xi2}) by including the full set of eigenstates of the $N_t=3$ sectors. 
In this appendix, we provide numerical estimates of both the parameter $R$, and its approximate value $\tilde R$, obtained by truncating the sums in Eqs. (\ref{Theta1} - \ref{Xi1}) to the two lowest excited states. 
Additionally, the estimate $\tilde{R}\approx 1$ obtained in Appendix \ref{app:Kondodet} relies on the strong interacting approximation in Eq. \eqref{Hevenred} for $\Delta_{\rm CAR} \gg E_C$.
\begin{figure*}[t!]
    \centering
    \includegraphics[width=\textwidth]{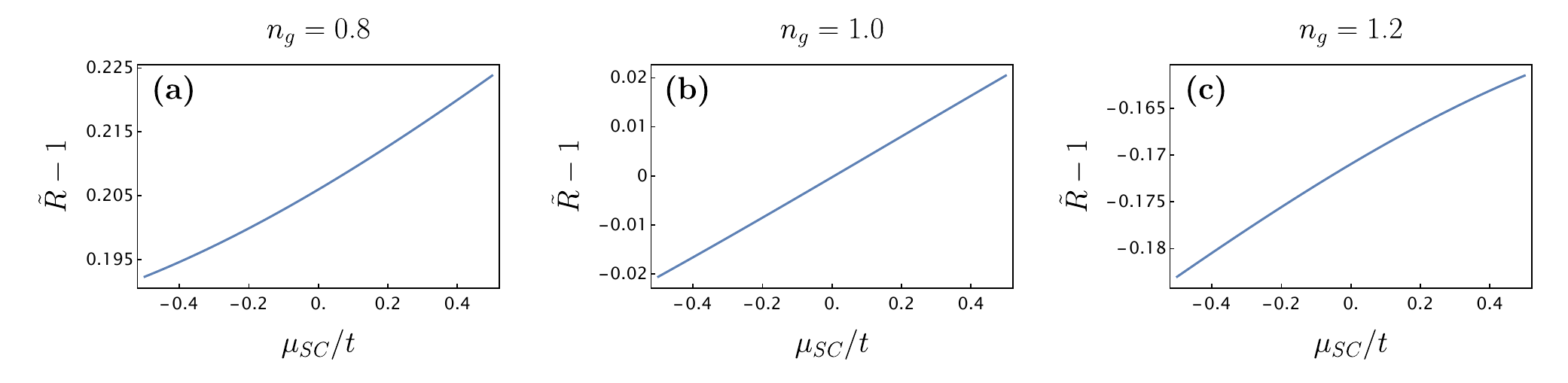}
    \caption{Ratio $\tilde R-1$ as a function of $\mu_{SC}$ for (a) $n_g=0.8$, (b) $n_g=1.0$ and (c) $n_g=1.2$. The ratio is computed on the degeneracy line and for $\Delta=4t, \; E_C=0.5t, \; \alpha=\pi/6$.}
    \label{fig:GammaRatio}
\end{figure*} 

 The truncated value $\tilde{R}$, however, can be estimated numerically independently of the assumption of strong interaction.  This is done in Fig. \ref{fig:ECplot}(a), where we depict both $R$ and $\tilde R$ as a function of the charging energy $E_C$ in the regime $E_C < t \ll \Delta$. The shaded region in the panel is excluded, since it corresponds to the condition $E_{C}<\Delta_{\text{CAR}}/6$, which violates the lower bound for $E_C$ to obtain a well-defined Kondo model, as discussed in Sec. \ref{sec:experiments}. Our results show that $R$ and $\tilde R$ remain close to one in a broad range of values of $E_C$, where the condition $E_C\gg\Delta_{\text{CAR}}$ is not necessarily verified. This confirms that our estimate $R\approx 1$ does not rely on the approximation in Eq. \eqref{Hevenred} and that it is robust against the inclusion of the contributions coming from the higher excited states in the $(3,\pm)$ sectors. This is evident also from Fig. \ref{fig:ECplot}(b), where we compute the exact ratio $R$ and compare it to the approximate value $\tilde R$ as a function of $\mu_{SC}$. We find that the contribution of the excited states amounts to a correction of order $\sim 2\%$ on the ratio $\tilde R$, over the range of parameters shown in the plot, which confirms that the truncation to the lowest energy states already provides an accurate estimate of $R$ for a broad range of the potential $\mu_{SC}$.
 
 Finally, in Fig. \ref{fig:GammaRatio}, we observe that $\tilde R$ is robust against changes in the potential $\mu_{SC}$ and remains of order $1$ also for variations $\pm0.2$ of the induced charge $n_g$ around the charge-degeneracy point.  Indeed, in the range of parameters shown in the plot, we find $\tilde R\in\left(0.75,1.25\right)$. Our estimates of $\tilde R$ in Fig. \ref{fig:GammaRatio} have been obtained by diagonalizing the $4\times 4$ Hamiltonians in Appendix \ref{app:sectors}, and truncating the sums in Eqs. (\ref{Theta1} - \ref{Xi1}) by considering only the two lowest excited states.

\section{Conductance close to the fixed point}
\label{app:SC-conductance}

\subsection{Fixed point}

In this Appendix, we derive the fixed point Emery--Kivelson quadratic action in Eq. \eqref{EK-0D-action} and evaluate explicitly the current-current correlation function. Neglecting the decoupled $c$, $s$, and $st$
sectors, the effective Hamiltonian of our model at the Emery--Kivelson point is
\begin{multline}
H_{*,t}=-i v_F\int_{-\infty}^{+\infty}dx\,\psi_t^\dagger(x)\partial_x\psi_t(x)\\
+iJ^{(\perp)}\sqrt{\frac{a}{2\pi}}\left[\psi_t(0)+\psi_t^\dagger(0)\right]\gamma_1 ,
\label{app-EK-H}
\end{multline}
We decompose the chiral fermion $\psi_t(x)$ into the Majorana modes
\begin{equation}
\xi_t=\psi_t+\psi_t^\dagger,\qquad\eta_t=i\left(\psi_t^\dagger-\psi_t\right).
\label{app-lead-Majoranas}
\end{equation}
The boundary interaction in Eq. \eqref{app-EK-H} couples $\xi_t(0)$ to $\gamma_1$, whereas $\eta_t(0)$ remains free.

We now integrate out the gapless bulk modes while keeping their boundary values fixed. In the wide-band limit, the local propagators of the two boundary Majorana modes read
\begin{equation}
G_{\chi}^{(0)}(i\omega_n)=2\int\frac{dk}{2\pi}\,\frac{1}{i\omega_n-v_Fk}=-\frac{i}{v_F}\,{\rm sgn}(\omega_n),
\label{app-local-lead-propagator}
\end{equation}
where $\chi=\xi_t(0),\eta_t(0)$ and $\omega_n=(2n+1)\pi/\beta$ is the Matsubara frequency of the corresponding fermionic field
\begin{equation}
    \chi_{\omega_n}=\int_{0}^{\beta} d\tau\,\ee^{i\omega_n \tau}\chi(\tau)
\end{equation}
The inverse of Eq. \eqref{app-local-lead-propagator} gives the dissipative boundary kernel $\mathcal K_{\chi}^{(0)}(i\omega_n)=i v_F{\rm sgn}(\omega_n)$. The resulting local Euclidean action is therefore:
\begin{multline}
S_{\rm imp}= \\
\frac{1}{2\beta}\sum_{\omega_n}\Big[\xi_{t,-\omega_n}\mathcal K_{\xi}^{(0)}(i\omega_n)\xi_{t,\omega_n}+\eta_{t,-\omega_n}\mathcal K_{\eta}^{(0)}(i\omega_n)\eta_{t,\omega_n}\Big]\\
+\frac{1}{4\beta}\sum_{\omega_n}\Big[\gamma_{1,-\omega_n}(-i\omega_n)\gamma_{1,\omega_n}+\gamma_{2,-\omega_n}(-i\omega_n)\gamma_{2,\omega_n}\Big]\\
+\frac{iJ^{(\perp)}}{\beta}\sqrt{\frac{a}{2\pi}}\sum_{\omega_n}\xi_{t,-\omega_n}\gamma_{1,\omega_n}.
\label{app-local-action}
\end{multline}
Since $\xi_t$ enters quadratically, it can be integrated out exactly. It generates the self-energy for $\gamma_1$
\begin{equation}
\Sigma_{\gamma_1}(i\omega_n)=-i\Gamma\operatorname{sgn}(\omega_n),\qquad\Gamma=\left(J^{(\perp)}\right)^2\frac{a}{\pi v_F}.
\label{app-Gamma}
\end{equation}
The effective fixed-point action \eqref{app-local-action} becomes the quadratic action $S_*$ in Eq. \eqref{EK-0D-action}.

To evaluate the conductance, we introduce the current correlator in Matsubara frequency space:
\begin{equation}
\Pi_{II}(i\Omega_n)=\int_0^\beta d\tau\,\ee^{i\Omega_n\tau}\left\langle T_\tau I^{\rm A}(\tau)I^{\rm A}(0)\right\rangle_* ,
\label{app-current-correlator}
\end{equation}
where $\Omega_n=2\pi n/\beta$ is a bosonic Matsubara frequency and the subscript $*$ indicates that the expectation value is calculated based on the action $S_*$. Equation~\eqref{Kubo} can then be written as
\begin{equation}
G_{\rm CAR}=\lim_{\omega\rightarrow0}\frac{1}{\omega}{\rm Im}\left[\left.\Pi_{II}(i\Omega_n)\right\vert_{\,i\Omega_n\rightarrow\omega+i0^+}\right].
\label{app-Kubo-frequency}
\end{equation}
 We shall therefore perform the calculation entirely in Matsubara-frequency space and analytically continue only the final result.

The fermionic propagators following from Eq. \eqref{EK-0D-action} are defined as
\begin{equation}
G_{\chi}(i\omega_n)=\int_0^\beta d\tau\,\ee^{i\omega_n\tau}\left\langle T_\tau\chi(\tau)\chi(0)\right\rangle_* ,
\end{equation}
and explicitly read
\begin{align}
G_{\eta_t}(i\omega_n)&=\frac{1}{i v_F\operatorname{sgn}(\omega_n)}=-\frac{i}{v_F}\operatorname{sgn}(\omega_n),
\label{app-eta-propagator}
\\
G_{\gamma_1}(i\omega_n)&=-\frac{2i\operatorname{sgn}(\omega_n)}{\left| \omega_n \right| +\Gamma},
\label{app-gamma1-propagator}
\\
G_{\gamma_2}(i\omega_n)&=-\frac{2i}{\omega_n}.
\label{app-gamma2-propagator}
\end{align}
Here $\omega_n=(2n+1)\pi/\beta$. The decoupled Majorana $\gamma_2$ does not enter the fixed-point current correlator, but will be required when the leading irrelevant interaction is included.
Using Eq.~\eqref{app-lead-Majoranas}, the current operator
\eqref{EK-current-refermionized} becomes $I^{\rm A}(\tau)=-i e \sqrt{a/2\pi}J^{(\perp)}\,\eta_t(\tau)\gamma_1(\tau)$. Wick's theorem therefore gives
\begin{equation}
\left\langle T_\tau I^{\rm A}(\tau)I^{\rm A}(0)\right\rangle_*=e^2\frac{a}{2\pi}\left(J^{(\perp)}\right)^2\,G_{\eta_t}(\tau)G_{\gamma_1}(\tau),
\end{equation}
and its Fourier transform is
\begin{equation}
\Pi_{II}^{*}(i\Omega_n)=\frac{e^2a\left(J^{(\perp)}\right)^2}{2\pi\beta}\sum_{\omega_m}G_{\eta_t}(i\omega_m)G_{\gamma_1}(i\Omega_n-i\omega_m).
\label{app-fixed-point-bubble}
\end{equation}

For $|\omega_m|,|\Omega_n|\ll\Gamma$, the propagators reduce to
\begin{equation}
G_{\eta_t}(i\omega_m)\simeq-\frac{i}{v_F}{\rm sgn}(\omega_m),\qquad G_{\gamma_1}(i\omega_m)\simeq-\frac{2i}{\Gamma}{\rm sgn}(\omega_m).
\end{equation}
Using $\Gamma=(J^{(\perp)})^2a/(\pi v_F)$, Eq.~\eqref{app-fixed-point-bubble} becomes
\begin{equation}
\Pi_{II}^{*}(i\Omega_n)=-\frac{e^2}{\beta}\sum_{\omega_m}{\rm sgn}(\omega_m)\,{\rm sgn}(\Omega_n-\omega_m).
\end{equation}
For the Kubo formula \eqref{app-Kubo-frequency}, only the linear frequency contribution of Eq. \eqref{app-second-order-triple-sum} under the analytical continuation $i\Omega_n\to\omega+i0^+$ in Eq.\eqref{app-fixed-point-bubble} will contribute to the zero frequency conductance. 
We then subtract the static contribution,
\begin{equation}
\Delta\Pi_{II}^{*}(i\Omega_n)\equiv\Pi_{II}^{*}(i\Omega_n)-\Pi_{II}^{*}(0).
\end{equation}
For $\Omega_n=2\pi n/\beta>0$, only the $n$ fermionic frequencies $0<\omega_m<\Omega_n$ contribute to the difference, yielding
\begin{equation}
\Delta\Pi_{II}^{*}(i\Omega_n)=-\frac{2e^2n}{\beta}=-\frac{e^2}{\pi}\Omega_n .
\end{equation}
Including both signs of $\Omega_n$,
\begin{equation}
\Delta\Pi_{II}^{*}(i\Omega_n)=-\frac{e^2}{\pi}|\Omega_n|.
\label{app-fixed-point-correlator}
\end{equation}
The retarded correlator is obtained by continuing the positive-frequency branch,
\begin{equation}
|\Omega_n|\longrightarrow-i(\omega+i0^+),
\end{equation}
so that the Kubo formula \eqref{app-Kubo-frequency} gives
\begin{equation}
G_{\rm CAR}^{*}=\frac{e^2}{\pi}=\frac{2e^2}{h},
\label{app-fixed-point-conductance}
\end{equation}
where we have reestablished the correct standard units in the last equality.

\subsection{Perturbation of the fixed point}
We now evaluate the leading correction generated by the residual longitudinal interaction in Eq. \eqref{EK-irrelevant-operator}. The first non-vanishing contribution is second order in $\lambda_z$:
\begin{multline}
\Pi_{II}^{(2)}(i\Omega_n)=\frac{\pi^2\lambda_z^2}{2}\int_0^\beta d\tau\,d\tau_1\,d\tau_2\,\ee^{i\Omega_n\tau}
\\
\times\left\langle T_\tau I^{\rm A}(\tau)I^{\rm A}(0)\mathcal O_{3/2}(\tau_1)\mathcal O_{3/2}(\tau_2)\right\rangle_{*,{\rm c}},
\label{app-second-order-time}
\end{multline}
when the subscript ${\rm c}$ denotes the connected part. Using the definition of the current, the connected Wick contractions factorize into the $\eta_t$, $\gamma_1$, $\gamma_2$, and charge-fermion sectors. By defining 
\[
C_c(\tau)=\left\langle T_\tau:\!\psi_c^\dagger\psi_c\!:(\tau):\!\psi_c^\dagger\psi_c\!:(0)\right\rangle_* ,
\]
one obtains
\begin{multline}
\Pi_{II}^{(2)}(i\Omega_n)=\frac{e^2\pi a\left(J^{(\perp)}\right)^2\lambda_z^2}{4}\int_0^\beta d\tau\,d\tau_1\,d\tau_2\,e^{i\Omega_n\tau}\\
\times G_{\eta_t}(\tau)C_c(\tau_1-\tau_2)G_{\gamma_2}(\tau_1-\tau_2)
\\
\times\left[-G_{\gamma_1}(\tau-\tau_1)G_{\gamma_1}(-\tau_2)+G_{\gamma_1}(\tau-\tau_2)G_{\gamma_1}(-\tau_1)\right].
\label{app-second-order-wick}
\end{multline}
The two terms in the square brackets give the same contribution after the exchange $\tau_1\leftrightarrow\tau_2$.

We evaluate Eq. \eqref{app-second-order-wick} in frequency space. Besides the propagators in Eqs. (\ref{app-eta-propagator},\ref{app-gamma1-propagator},\ref{app-gamma2-propagator}), we need the boundary charge-fermion propagator
\begin{equation}
G_{\psi_c}(i\omega_m)=-\frac{i}{2v_F}\operatorname{sgn}(\omega_m)\,.
\label{app-psic-correlator}
\end{equation}
Fourier transforming all the correlators in Eq.~\eqref{app-second-order-wick} gives
\begin{multline}
\Pi_{II}^{(2)}(i\Omega_n)=
\frac{e^2\pi^2 v_F \Gamma\lambda_z^2}
{2\beta }\\
\times
\sum_{\omega_m}G_{\eta}(i\Omega_n-i\omega_m)K(i\omega_m) G_{\gamma_1}(i\omega_m) G_{\gamma_1}(-i\omega_m)
\label{app-second-order-triple-sum}
\end{multline}
where
\begin{equation}
K(i\omega_m)=\frac{1}{\beta^2}\sum_{\omega_l, \omega_p}G_{\psi_c}(i\omega_l) G_{\psi_c}(i\omega_p) G_{\gamma_2}(i\omega_m-i\omega_l +i \omega_p),
\label{app-K-product}
\end{equation}
and we have used the definition of $\Gamma=\left(J^{(\perp)}\right)^2 a/(\pi v_F)$.
 All sums are over fermionic Matsubara frequencies and are understood with a hard ultraviolet cutoff $N_D\sim D\beta/(2\pi)$ over the Matsubara indices $l,p$, inherited from the bandwidth $D$ of the integrated out bulk modes.

For $N_D\gg m$, the internal Matsubara sums give
\begin{multline}
K(i\omega_m)=\frac{i}{2v_F^2}\frac{(2m+1)}{2\pi\beta}\left[2\log N_D+A_m\right],
\label{app-KN-digamma}
\end{multline}
where
\begin{equation}
A_m=-\psi\!\left(\frac12-m\right)-\psi\!\left(m+\frac32\right).
\label{app-Am-result}
\end{equation}
Here $\psi(z)=d\log\Upgamma(z)/dz$ is the digamma function.

For $\Omega_n=2\pi n/\beta>0$, only the $n$ fermionic frequencies $0<\omega_m<\Omega_n$ contribute to the difference $\Delta\Pi^{(2)}_{II}(i\Omega_n)=\Pi^{(2)}_{II}(i\Omega_n)-\Pi_{II}^{(2)}(0)$. In the regime $\beta^{-1}, |\Omega_n|\ll\Gamma$, we obtain
\begin{equation}
\Delta\Pi_{II}^{(2)}(i\Omega_n)=
\frac{e^2\pi\lambda_z^2}{v_F^2\beta^2\Gamma}\sum_{m=0}^{n-1}\left(2m+1\right)\left(2\ln N_D+A_m\right).
\label{app-DPi-second-before-sum}
\end{equation}
Using the reflection and recurrence relations of the digamma function \cite{Erdelyi1953}, the finite sum in Eq. \eqref{app-DPi-second-before-sum} can then be performed explicitly. One finds
\begin{equation}
\Delta\Pi_{II}^{(2)}(i\Omega_n)=\frac{e^2\pi^2\lambda_z^2}{8v_F^2}\frac{T}{\Gamma}|\Omega_n|+O(\Omega_n^2),
\label{app-DPi-second-final}
\end{equation}
Finally, by analytic continuation of the positive Matsubara branch and adding the fixed point result, we obtain
\begin{equation}
G_{\rm CAR}(T)=\frac{2e^2}{h}\left[
1-\frac{\pi^3\lambda_z^2}{8v_F^2}\frac{T}{\Gamma}+O\left(\frac{T^2}{\Gamma^2}\right)\right].
\label{app-GCAR-correction-standard}
\end{equation}

\bibliography{MPStransport}
\end{document}